\pdfoutput=1

\documentclass[12pt,a4paper]{article}

\usepackage{ifthen} 
\newboolean{pdflatex}
\setboolean{pdflatex}{true} 

\newboolean{articletitles}
\setboolean{articletitles}{true} 

\newboolean{uprightparticles}
\setboolean{uprightparticles}{false} 

\usepackage{microtype}
\usepackage{lineno}  
\usepackage{xspace} 

\usepackage{graphicx}  
\usepackage{color}
\usepackage{colortbl}
\graphicspath{{./figs/}} 
\usepackage{pinlabel}
\usepackage{multirow}

\usepackage{amsmath} 
\usepackage{amssymb}
\usepackage{amsfonts}
\usepackage{upgreek} 

\newcommand*\patchAmsMathEnvironmentForLineno[1]{%
\expandafter\let\csname old#1\expandafter\endcsname\csname #1\endcsname
\expandafter\let\csname oldend#1\expandafter\endcsname\csname
end#1\endcsname
 \renewenvironment{#1}%
   {\linenomath\csname old#1\endcsname}%
   {\csname oldend#1\endcsname\endlinenomath}%
}
\newcommand*\patchBothAmsMathEnvironmentsForLineno[1]{%
  \patchAmsMathEnvironmentForLineno{#1}%
  \patchAmsMathEnvironmentForLineno{#1*}%
}
\AtBeginDocument{%
\patchBothAmsMathEnvironmentsForLineno{equation}%
\patchBothAmsMathEnvironmentsForLineno{align}%
\patchBothAmsMathEnvironmentsForLineno{flalign}%
\patchBothAmsMathEnvironmentsForLineno{alignat}%
\patchBothAmsMathEnvironmentsForLineno{gather}%
\patchBothAmsMathEnvironmentsForLineno{multline}%
}

\usepackage{hyperref}    

\def\Dbp    {\ensuremath{\Dp_{(\squark)}}\xspace}					
\def\Mmumu   {\ensuremath{m(\mumu)}\xspace}					
\def\Mpimu   {\ensuremath{m(\pim\mup)}\xspace}					

\def\Dpmmos  {\ensuremath{\decay{\Dp}{\pip\mumu}}\xspace}			
\def\Dsmmos  {\ensuremath{\decay{\Ds}{\pip\mumu}}\xspace}			
\def\Dbmmos    {\ensuremath{\decay{\Dbp}{\pip\mumu}}\xspace}		
\def\Dpmmss    {\ensuremath{\decay{\Dp}{\pim\mup\mup}}\xspace}		
\def\Dsmmss    {\ensuremath{\decay{\Ds}{\pim\mup\mup}}\xspace}		
\def\Dbmmss    {\ensuremath{\decay{\Dbp}{\pim\mup\mup}}\xspace}		
\def\Dpmmns    {\ensuremath{\decay{\Dbp}{\pi\mmu\mmu}}\xspace}		
\def\Dppp		{\ensuremath{\decay{\Dbp}{\pip\pip\pim}}\xspace}		  
\def\Dphipi	{\ensuremath{\decay{\Dbp}{\pip(\Pphi\to\mumu)}}\xspace} 
\def\Dpetapi	{\ensuremath{\decay{\Dp}{\pip(\Peta\to\mumu)}}\xspace}   
\def\Dsetapi	{\ensuremath{\decay{\Ds}{\pip(\Peta\to\mumu)}}\xspace}   
\def\cls     {\ensuremath{\mathrm{CL}_{s}}\xspace}					 
\def\clb     {\ensuremath{\mathrm{CL}_{b}}\xspace}					 
\def\cl     {\ensuremath{\mathrm{CL}}\xspace}						 

\def\lhcb {\mbox{LHCb}\xspace}
\def\ux85 {\mbox{UX85}\xspace}

\def\babar  {\mbox{BaBar}\xspace}

\ifthenelse{\boolean{uprightparticles}}%
{

 \def\Peta        {\ensuremath{\upeta}\xspace}

 \def\Pmu         {\ensuremath{\upmu}\xspace}

 \def\Ppi         {\ensuremath{\uppi}\xspace}                 
                  
 \def\Prho        {\ensuremath{\uprho}\xspace}

 \def\Pphi        {\ensuremath{\upphi}\xspace}

 \def\Pomega      {\ensuremath{\upomega}\xspace}                 

 \def\PDelta      {\ensuremath{\Delta}\xspace}                 
 \def\PXi      {\ensuremath{\Xi}\xspace}                 
 \def\PLambda      {\ensuremath{\Lambda}\xspace}                 
 \def\PSigma      {\ensuremath{\Sigma}\xspace}                 
 \def\POmega      {\ensuremath{\Omega}\xspace}                 
 \def\PUpsilon      {\ensuremath{\Upsilon}\xspace}                 
 
 \def\PB      {\ensuremath{\mathrm{B}}\xspace}                 
                  
 \def\PD      {\ensuremath{\mathrm{D}}\xspace}

 \def\PK      {\ensuremath{\mathrm{K}}\xspace}

 \def\Pb      {\ensuremath{\mathrm{b}}\xspace}                 
 \def\Pc      {\ensuremath{\mathrm{c}}\xspace}                 
 \def\Pd      {\ensuremath{\mathrm{d}}\xspace}

 \def\Pi      {\ensuremath{\mathrm{i}}\xspace}

 \def\Ps      {\ensuremath{\mathrm{s}}\xspace}                 
                  
 \def\Pu      {\ensuremath{\mathrm{u}}\xspace}

}
{

 \def\Peta        {\ensuremath{\eta}\xspace}

 \def\Pmu         {\ensuremath{\mu}\xspace}

 \def\Ppi         {\ensuremath{\pi}\xspace}                 
                  
 \def\Prho        {\ensuremath{\rho}\xspace}

 \def\Pphi        {\ensuremath{\phi}\xspace}

 \def\Pomega      {\ensuremath{\omega}\xspace}                 
 \mathchardef\PDelta="7101
 \mathchardef\PXi="7104
 \mathchardef\PLambda="7103
 \mathchardef\PSigma="7106
 \mathchardef\POmega="710A
 \mathchardef\PUpsilon="7107
                  
 \def\PB      {\ensuremath{B}\xspace}                 
                  
 \def\PD      {\ensuremath{D}\xspace}

 \def\PK      {\ensuremath{K}\xspace}

 \def\Pb      {\ensuremath{b}\xspace}                 
 \def\Pc      {\ensuremath{c}\xspace}                 
 \def\Pd      {\ensuremath{d}\xspace}

 \def\Pi      {\ensuremath{i}\xspace}

 \def\Ps      {\ensuremath{s}\xspace}                 
                  
 \def\Pu      {\ensuremath{u}\xspace}

}

\def\mmu       {\ensuremath{\Pmu}\xspace}
\def\mup        {\ensuremath{\Pmu^+}\xspace}
\def\mumu     {\ensuremath{\Pmu^+\Pmu^-}\xspace}

\def\uquark    {\ensuremath{\Pu}\xspace}

\def\dquark    {\ensuremath{\Pd}\xspace}

\def\squark    {\ensuremath{\Ps}\xspace}

\def\cquark    {\ensuremath{\Pc}\xspace}

\def\bquark    {\ensuremath{\Pb}\xspace}

\def\pion  {\ensuremath{\Ppi}\xspace}

\def\pip   {\ensuremath{\pion^+}\xspace}
\def\pim   {\ensuremath{\pion^-}\xspace}

\def\kaon  {\ensuremath{K}\xspace}
  \def\Kbar  {\kern 0.2em\overline{\kern -0.2em \PK}{}\xspace}

\def\Kz    {\ensuremath{\kaon^0}\xspace}
\def\Kzb   {\ensuremath{\Kbar^0}\xspace}
\def\KzKzb {\ensuremath{\Kz \kern -0.16em \Kzb}\xspace}
\def\Kp    {\ensuremath{\kaon^+}\xspace}
\def\Km    {\ensuremath{\kaon^-}\xspace}

\def\KpKm  {\ensuremath{\Kp \kern -0.16em \Km}\xspace}

\def\Dbar    {\kern 0.2em\overline{\kern -0.2em \PD}{}\xspace}
\def\D       {\ensuremath{D}\xspace}

\def\Dz      {\ensuremath{\D^0}\xspace}
\def\Dzb     {\ensuremath{\Dbar^0}\xspace}
\def\DzDzb   {\ensuremath{\Dz {\kern -0.16em \Dzb}}\xspace}
\def\Dp      {\ensuremath{\D^+}\xspace}
\def\Dm      {\ensuremath{\D^-}\xspace}

\def\DpDm    {\ensuremath{\Dp {\kern -0.16em \Dm}}\xspace}

\def\Dstarp  {\ensuremath{\D^{*+}}\xspace}

\def\Ds      {\ensuremath{\D^+_\squark}\xspace}

\def\B       {\ensuremath{B}\xspace}
  \def\Bbar    {\kern 0.18em\overline{\kern -0.18em \PB}{}\xspace}

\def\Bu      {\ensuremath{\B^+}\xspace}

\def\Bp      {\ensuremath{\Bu}\xspace}

  \def\Y#1S{\ensuremath{\PUpsilon{(#1S)}}\xspace}

\def\Lbar {\ensuremath{\kern 0.1em\overline{\kern -0.1em\PLambda}}\xspace}

\def\BF         {{\ensuremath{\cal B}\xspace}}

\newcommand{\decay}[2]{\ensuremath{#1\!\to #2}\xspace}         

\def\to                 {\ensuremath{\rightarrow}\xspace}

\def\Vcd  {\ensuremath{|V_{\cquark\dquark}|}\xspace}

\def\AT#1     {\ensuremath{A_{\mathrm{T}}^{#1}}\xspace}           

\def\C#1      {\ensuremath{\mathcal{C}_{#1}}\xspace}                       
\def\Cp#1     {\ensuremath{\mathcal{C}_{#1}^{'}}\xspace}                    
\def\Ceff#1   {\ensuremath{\mathcal{C}_{#1}^{\mathrm{(eff)}}}\xspace}        
\def\Cpeff#1  {\ensuremath{\mathcal{C}_{#1}^{'\mathrm{(eff)}}}\xspace}       
\def\Ope#1    {\ensuremath{\mathcal{O}_{#1}}\xspace}                       
\def\Opep#1   {\ensuremath{\mathcal{O}_{#1}^{'}}\xspace}                    

\newcommand{\tev}{\ensuremath{\mathrm{\,Te\kern -0.1em V}}\xspace}
\newcommand{\gev}{\ensuremath{\mathrm{\,Ge\kern -0.1em V}}\xspace}
\newcommand{\mev}{\ensuremath{\mathrm{\,Me\kern -0.1em V}}\xspace}
\newcommand{\kev}{\ensuremath{\mathrm{\,ke\kern -0.1em V}}\xspace}
\newcommand{\ev}{\ensuremath{\mathrm{\,e\kern -0.1em V}}\xspace}
\newcommand{\gevc}{\ensuremath{{\mathrm{\,Ge\kern -0.1em V\!/}c}}\xspace}
\newcommand{\mevc}{\ensuremath{{\mathrm{\,Me\kern -0.1em V\!/}c}}\xspace}
\newcommand{\gevcc}{\ensuremath{{\mathrm{\,Ge\kern -0.1em V\!/}c^2}}\xspace}
\newcommand{\gevgevcccc}{\ensuremath{{\mathrm{\,Ge\kern -0.1em V^2\!/}c^4}}\xspace}
\newcommand{\mevcc}{\ensuremath{{\mathrm{\,Me\kern -0.1em V\!/}c^2}}\xspace}

\def\mum  {\ensuremath{\,\upmu\rm m}\xspace}

\def\invpb {\ensuremath{\mbox{\,pb}^{-1}}\xspace}

\def\invfb   {\ensuremath{\mbox{\,fb}^{-1}}\xspace}

\newcommand{\chisq}{\ensuremath{\chi^2}\xspace}

\def\gsim{{~\raise.15em\hbox{$>$}\kern-.85em
          \lower.35em\hbox{$\sim$}~}\xspace}
\def\lsim{{~\raise.15em\hbox{$<$}\kern-.85em
          \lower.35em\hbox{$\sim$}~}\xspace}

\def\ptot       {\mbox{$p$}\xspace}
\def\pt         {\mbox{$p_{\rm T}$}\xspace}

\def\dllkpi     {\ensuremath{\mathrm{DLL}_{\kaon\pion}}\xspace}

\def\dllmupi    {\ensuremath{\mathrm{DLL}_{\mmu\pi}}\xspace}

\def\degrees{\ensuremath{^{\circ}}\xspace}

\def\evtgen     {\mbox{\textsc{EvtGen}}\xspace}
\def\pythia     {\mbox{\textsc{Pythia}}\xspace}

\def\geant      {\mbox{\textsc{Geant4}}\xspace}

\def\tell1  {TELL1\xspace}
\def\ukl1   {UKL1\xspace}

\usepackage{tikz}
\usetikzlibrary{decorations.pathreplacing,decorations.pathmorphing,decorations.markings,trees,calc,patterns,snakes}
\usepackage{hepnames}
\usepackage{cleveref}
\usepackage{cite} 
\usepackage{mciteplus}
\mciteErrorOnUnknownfalse

\begin{document}

\renewcommand{\thefootnote}{\fnsymbol{footnote}}
\setcounter{footnote}{1}


\begin{titlepage}
\pagenumbering{roman}

\vspace*{-1.5cm}
\centerline{\large EUROPEAN ORGANIZATION FOR NUCLEAR RESEARCH (CERN)}
\vspace*{1.5cm}
\hspace*{-0.5cm}
\begin{tabular*}{\linewidth}{lc@{\extracolsep{\fill}}r}
\ifthenelse{\boolean{pdflatex}}
{\vspace*{-2.7cm}\mbox{\!\!\!\includegraphics[width=.14\textwidth]{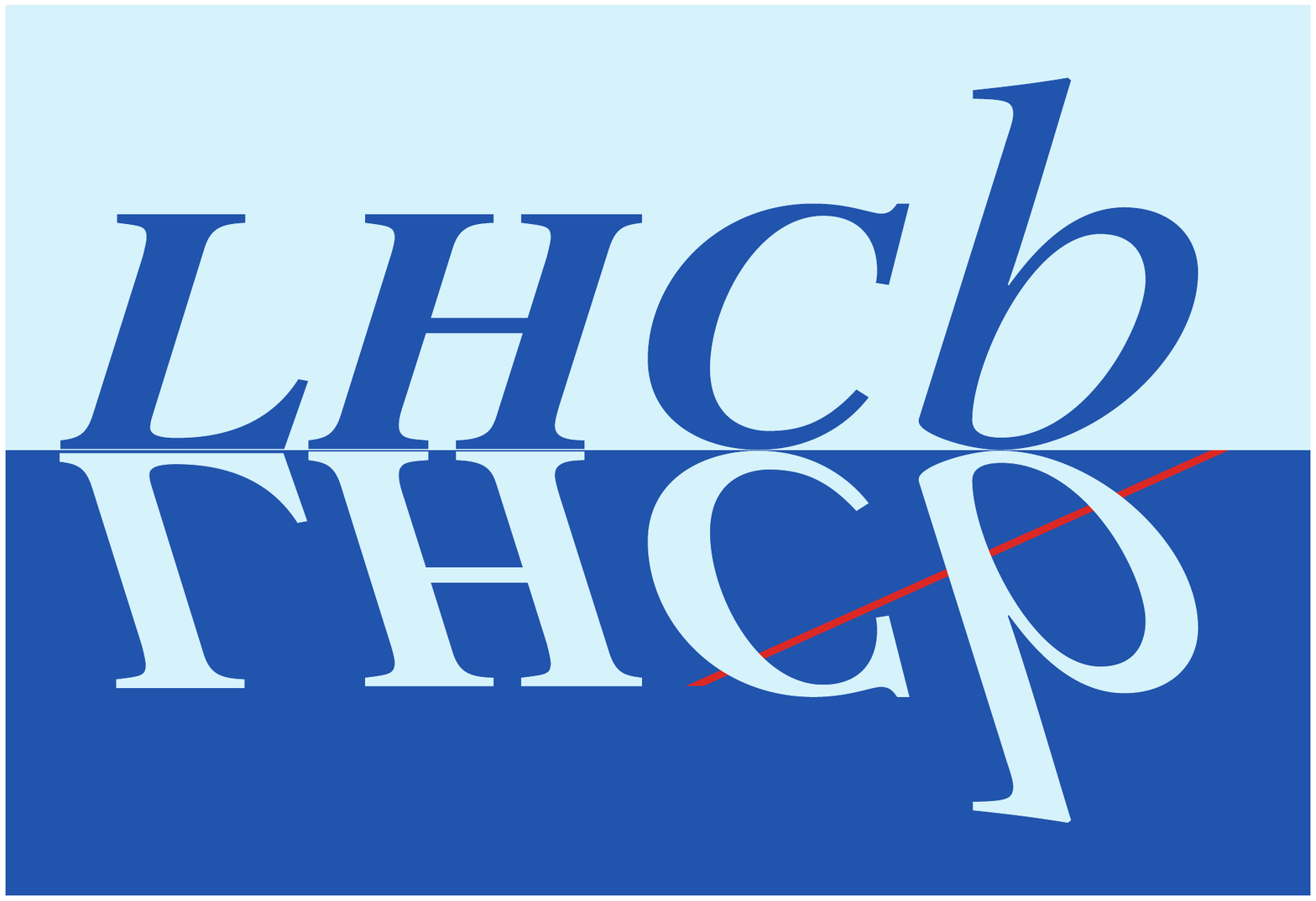}} & &}%
{\vspace*{-1.2cm}\mbox{\!\!\!\includegraphics[width=.12\textwidth]{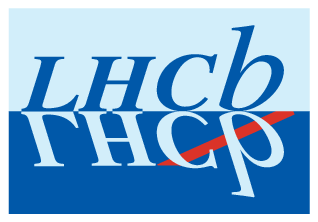}} & &}%
\\
 & & CERN-PH-EP-2013-061 \\  
 & & LHCb-PAPER-2012-051 \\  
 & & 23 April 2013 
\end{tabular*}

\vspace*{2.0cm}

{\bf\boldmath\huge
\begin{center}
  Search for \Dbmmos\\ and \Dbmmss decays
\end{center}
}

\vspace*{1.0cm}

\begin{center}
The LHCb collaboration\footnote{Authors are listed on the following pages.}
\end{center}

\vspace{\fill}

\begin{abstract}
\noindent
A search for non-resonant \Dbmmos and \Dbmmss decays is performed using proton-proton collision data, corresponding to an integrated luminosity of 1.0 \invfb, at $\sqrt{s}=7$ \tev recorded by the LHCb experiment in 2011.
No signals are observed and the $90\% \, (95\%)$ confidence level (\cl) limits on the branching fractions are found to be
\begin{eqnarray*}
\mathcal{\BF}(\Dpmmos) < 7.3 \, (8.3) \times 10^{-8},\\
\mathcal{\BF}(\Dsmmos) < 4.1 \, (4.8) \times 10^{-7},\\
\mathcal{\BF}(\Dpmmss) < 2.2 \, (2.5) \times 10^{-8},\\
\mathcal{\BF}(\Dsmmss) < 1.2 \, (1.4) \times 10^{-7}.\\
\end{eqnarray*}
These limits are the most stringent to date.
\end{abstract}
\vspace*{1.0cm}

\begin{center}
Submitted to Phys. Lett. B
\end{center}

\vspace{\fill}

{\footnotesize 
\centerline{\copyright~CERN on behalf of the \lhcb collaboration, license \href{http://creativecommons.org/licenses/by/3.0/}{CC-BY-3.0}.}}
\vspace*{2mm}

\end{titlepage}


\newpage
\setcounter{page}{2}
\mbox{~}
\newpage
%
\centerline{\large\bf LHCb collaboration}
\begin{flushleft}
\small
R.~Aaij$^{40}$, 
C.~Abellan~Beteta$^{35,n}$, 
B.~Adeva$^{36}$, 
M.~Adinolfi$^{45}$, 
C.~Adrover$^{6}$, 
A.~Affolder$^{51}$, 
Z.~Ajaltouni$^{5}$, 
J.~Albrecht$^{9}$, 
F.~Alessio$^{37}$, 
M.~Alexander$^{50}$
% ...

{\footnotesize \it
% ...
$ ^{5}$Clermont Universit\'{e}, Universit\'{e} Blaise Pascal, CNRS/IN2P3, LPC, Clermont-Ferrand, France\\
$ ^{6}$CPPM, Aix-Marseille Universit\'{e}, CNRS/IN2P3, Marseille, France\\
% ...
$ ^{9}$Fakult\"{a}t Physik, Technische Universit\"{a}t Dortmund, Dortmund, Germany\\
% ...
$ ^{35}$Universitat de Barcelona, Barcelona, Spain\\
$ ^{36}$Universidad de Santiago de Compostela, Santiago de Compostela, Spain\\
$ ^{37}$European Organization for Nuclear Research (CERN), Geneva, Switzerland\\
% ...
$ ^{40}$Nikhef National Institute for Subatomic Physics, Amsterdam, The Netherlands\\
% ...
$ ^{45}$H.H. Wills Physics Laboratory, University of Bristol, Bristol, United Kingdom\\
% ...
$ ^{50}$School of Physics and Astronomy, University of Glasgow, Glasgow, United Kingdom\\
$ ^{51}$Oliver Lodge Laboratory, University of Liverpool, Liverpool, United Kingdom\\
% ...
$ ^{n}$LIFAELS, La Salle, Universitat Ramon Llull, Barcelona, Spain\\
% ...
}
\end{flushleft}
%
\cleardoublepage


\renewcommand{\thefootnote}{\arabic{footnote}}
\setcounter{footnote}{0}



\pagestyle{plain} 
\setcounter{page}{1}
\pagenumbering{arabic}


%

\tikzset{
photon/.style={decorate, decoration={snake}, draw=red},
particle/.style={draw=blue, postaction={decorate},decoration={markings,mark=at position .5 with {\arrow[draw=blue]{>}}}},
antiparticle/.style={draw=blue, postaction={decorate},decoration={markings,mark=at position .5 with {\arrow[draw=blue]{<}}}}, 
gluon/.style={decorate, draw=black,decoration={snake,amplitude=4pt, segment length=5pt}}, 
majorana/.style={draw=black, postaction={decorate},decoration={markings,mark=at position .48 with {\arrow[draw=black]{>}},mark=at position .52 with {\arrow[draw=black]{<}}}},
gluonloop/.style={circle, decorate, draw=black, decoration={coil,aspect=1.2,amplitude=2pt, segment length=4pt},minimum height=1.2em},
}

\section{Introduction}
\label{sec:Introduction}

\begin{figure}[!h]
\centering

\begin{tikzpicture}[scale=1.0]
\begin{scope}
\coordinate (a) at (0,1); 
\coordinate (b) at (0,0); 
\coordinate (c) at (4,1); 
\coordinate (d) at (4,0); 
\coordinate (e) at (1,1); 
\coordinate (f) at (2,1); 
\coordinate (g) at (3,1); 
\coordinate (h) at (3,2.5); 
\coordinate (i) at (4,2); 
\coordinate (j) at (4,3); 
\draw[particle] (a) -- (f);
\draw[particle] (f) -- (c);
\draw[antiparticle] (b) -- (d);
\draw[photon] (e) to [out=270,in=270]  (g) ; 
\draw[photon] (f) -- (h);
\draw[particle] (h) -- (i);
\draw[antiparticle] (h) -- (j);
\node at ($(a)$) [label={[label distance=10mm] above:(a)}] {};
\node at ($(a)$) [label={[label distance=-1.5mm] left:$\Pqc$}] {};
\node at ($(b)$) [label={[label distance=-1.5mm] left:$\APqd$}] {};
\node at ($(c)$) [label={[label distance=-1.5mm] right:$\Pqu$}] {};
\node at ($(d)$) [label={[label distance=-1.5mm] right:$\APqd$}] {};
\node at ($(f)$) [label={[label distance=5mm] above:$\Pphoton/\PZz$}] {};
\node at ($(f)$) [label={[label distance=-1mm] below:$\PWp$}] {};	
\node at ($(i)$) [label={[label distance=-1.5mm] right:$\Pmuon$}] {};
\node at ($(j)$) [label={[label distance=-1.5mm] right:$\APmuon$}] {};
\draw [black,decorate,decoration={brace,amplitude=5pt},xshift=-17pt,yshift=0pt]
  (0,0)  -- (0,1) node [black,midway,left=0pt,xshift=-5pt] {$\Dp$};
\draw [black,decorate,decoration={brace,amplitude=5pt},xshift=17pt,yshift=0pt]
  (4,1)  -- (4,0) node [black,midway,right=0pt,xshift=5pt] {$\pip$};
\end{scope}

\begin{scope}[xshift=8 cm]
\coordinate (a) at (0,1); 
\coordinate (b) at (0,0); 
\coordinate (c) at (4,1); 
\coordinate (d) at (4,0); 
\coordinate (e) at (1,1); 
\coordinate (f) at (2,2); 
\coordinate (g) at (3,1); 
\coordinate (h) at (4,2); 
\coordinate (i) at (3,3); 
\coordinate (j) at (5,3); 
\draw[particle] (a) -- (c);
\draw[antiparticle] (b) -- (d);
\draw[photon] (e) -- (f);
\draw[photon] (g) -- (h);
\draw[antiparticle] (f) -- (i);
\draw[particle] (f) -- (h);
\draw[particle] (h) -- (j);
\node at ($(a)$) [label={[label distance=10mm] above:(b)}] {};
\node at ($(a)$) [label={[label distance=-1.5mm] left:$\Pqc$}] {};
\node at ($(b)$) [label={[label distance=-1.5mm] left:$\APqd$}] {};
\node at ($(c)$) [label={[label distance=-1.5mm] right:$\Pqu$}] {};
\node at ($(d)$) [label={[label distance=-1.5mm] right:$\APqd$}] {};
\node at ($(e)$) [label={[label distance=+1.5mm] above:$\PWp$}] {};
\node at ($(g)$) [label={[label distance=+1.5mm] above:$\PWm$}] {};	
\node at ($(i)$) [label={[label distance=-1.5mm] right:$\APmuon$}] {};
\node at ($(j)$) [label={[label distance=-1.5mm] right:$\Pmuon$}] {};
\draw [black,decorate,decoration={brace,amplitude=5pt},xshift=-17pt,yshift=0pt]
  (0,0)  -- (0,1) node [black,midway,left=0pt,xshift=-5pt] {$\Dp$};
\draw [black,decorate,decoration={brace,amplitude=5pt},xshift=17pt,yshift=0pt]
  (4,1)  -- (4,0) node [black,midway,right=0pt,xshift=5pt] {$\pip$};
\end{scope}
\end{tikzpicture}

\begin{tikzpicture}[scale=1.0]

\begin{scope}
\coordinate (a) at (0,1); 
\coordinate (b) at (0,0); 
\coordinate (c) at (1,0.5); 
\coordinate (d) at (3,0.5); 
\coordinate (k) at (2,0.5); 
\coordinate (e) at (4,1); 
\coordinate (f) at (4,0); 
\coordinate (h) at (3,2.5); 
\coordinate (i) at (4,2); 
\coordinate (j) at (4,3); 
\draw[particle] (a) -- (c);
\draw[antiparticle] (b) -- (c);
\draw[photon] (c) -- (d);
\draw[antiparticle] (d) -- (e);
\draw[particle] (d) -- (f);
\draw[photon] (k) -- (h);
\draw[particle] (h) -- (i);
\draw[antiparticle] (h) -- (j);
\node at ($(a)$) [label={[label distance=10mm] above:(c)}] {};
\node at ($(a)$) [label={[label distance=-1.5mm] left:c}] {};
\node at ($(b)$) [label={[label distance=-1.5mm] left:$\bar{\rm d}$,$\bar{\rm s}$}] {};
\node at ($(c)$) [label={[label distance=-1mm,xshift=10pt,] below right:$W^+$}] {};
\node at ($(e)$) [label={[label distance=-1.5mm] right:$\Pqu$}] {};
\node at ($(f)$) [label={[label distance=-1.5mm] right:$\APqd$}] {};
\node at ($(k)$) [label={[label distance=5mm] above:$\Pphoton/\PZz$}] {};
\node at ($(i)$) [label={[label distance=-1.5mm] right:$\mu^-$}] {};
\node at ($(j)$) [label={[label distance=-1.5mm] right:$\mu^+$}] {};
\draw [black,decorate,decoration={brace,amplitude=5pt},xshift=-17pt,yshift=0pt]
  (-0.2,0)  -- (-0.2,1) node [black,midway,left=0pt,xshift=-5pt] {$\Dbp$};
\draw [black,decorate,decoration={brace,amplitude=5pt},xshift=17pt,yshift=0pt]
  (4,1)  -- (4,0) node [black,midway,right=0pt,xshift=5pt] {$\pip$};
\end{scope}

\begin{scope}[xshift=7 cm]
\coordinate (a) at (0,3); 
\coordinate (b) at (0,2); 
\coordinate (c) at (1,2.5); 
\coordinate (d) at (2.5,2.5); 
\coordinate (e) at (3.5,3.5); 
\coordinate (f) at (3.5,1.5); 
\coordinate (g) at (4.5,2.5); 
\coordinate (h) at (4.5,0.5); 
\coordinate (i) at (5.5,1); 
\coordinate (j) at (5.5,0); 
\draw[particle] (a) -- (c);
\draw[antiparticle] (b) -- (c);
\draw[photon] (c) -- (d);
\draw[antiparticle] (d) -- (e);
\draw[majorana] (d) -- (f);
\draw[antiparticle] (f) -- (g);
\draw[photon] (f) -- (h);
\draw[particle] (h) -- (i);
\draw[particle] (j) -- (h);
\node at ($(c)$) [label={[label distance=+15mm] below:(d)}] {};
\node at ($(a)$) [label={[label distance=-1.5mm] left:c}] {};
\node at ($(b)$) [label={[label distance=-1.5mm] left:$\bar{\rm d}$,$\bar{\rm s}$}] {};
\node at ($(i)$) [label={[label distance=-1.5mm] right:$\bar{\rm u}$}] {};
\node at ($(j)$) [label={[label distance=-1.5mm] right:d}] {};
\node at ($(c)$) [label={[label distance=1mm,xshift=10pt,] above right:$W^+$}] {};
\node at ($(f)$) [label={[label distance=3mm] below:$W^-$}] {}; 
\node at ($(e)$) [label={[label distance=-1.5mm] right:$\mu^+$}] {};
\node at ($(g)$) [label={[label distance=-1.5mm] right:$\mu^+$}] {};
\node at ($(d)$) [label={[label distance=3mm] below :$\nu$}] {};
\draw [black,decorate,decoration={brace,amplitude=5pt},xshift=-17pt,yshift=0pt]
 (-0.2,2)  -- (-0.2,3) node [black,midway,left=0pt,xshift=-5pt] {$D_{(s)}^+$};
\draw [black,decorate,decoration={brace,amplitude=5pt},xshift=17pt,yshift=0pt]
 (5.5,1)  -- (5.5,0) node [black,midway,right=0pt,xshift=5pt] {$\pi^-$};
 \end{scope}
  
\end{tikzpicture}
\caption{\small{Feynman diagrams for (a,b) the FCNC decay \Dpmmos, (c) the weak annihilation of a \Dbp meson and (d) a possible LNV \Dbp meson decay mediated by a Majorana neutrino.}}
\label{fig:FDs}
\end{figure}
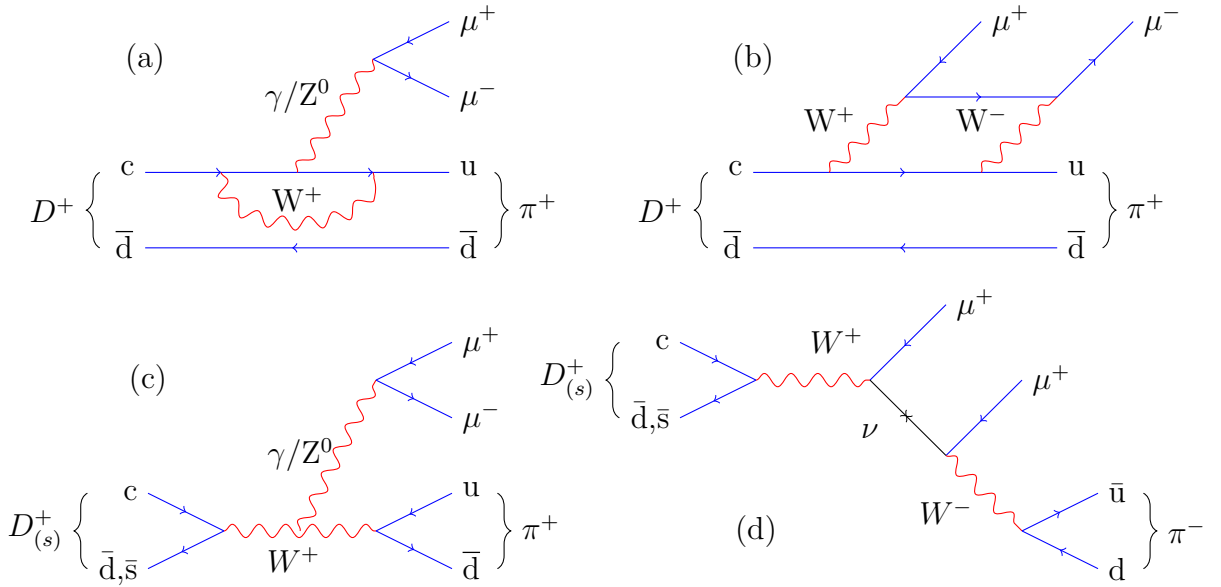

Flavour-changing neutral current (FCNC) processes are rare within the Standard Model (SM) as they cannot occur at tree level. 
At the loop level, they are suppressed by the GIM mechanism~\cite{Fajfer:2006yc} but are nevertheless well established in $\Bp\to\Kp\mumu$ and $\Kp\to\pip\mumu$ decays with branching fractions of the order $10^{-7}$ and $10^{-8}$, respectively~\cite{Abe:2001dh,Park:2001cv}. In contrast to the \B meson system, where the very high mass of the top quark in the loop weakens the suppression, the GIM cancellation is almost exact in \D meson decays leading to expected branching fractions for \cquark\to\uquark\mumu processes in the $(1-3)\times10^{-9}$ range~\cite{Fajfer:2001sa,Fajfer:2007dy,Paul:2011ar}. This suppression provides a unique opportunity to search for FCNC \D meson decays and to probe the coupling of up-type quarks in electroweak processes, as illustrated in Fig.~\ref{fig:FDs}(a,b).

The decay \Dsmmos, although not a FCNC process, proceeds via the weak annihilation diagram shown in Fig.~\ref{fig:FDs}(c). This can be used to normalise a potential \Dpmmos signal where an analogous weak annihilation diagram proceeds, albeit suppressed by a factor $\Vcd^2$. Normalisation is needed in order to distinguish between FCNC and weak annihilation contributions. Note that, throughout this paper, the inclusion of conjugate processes is implied.

Many extensions of the SM, such as Supersymmetric models with R-parity violation or models involving a fourth quark generation, introduce additional diagrams that \emph{a priori} need not be suppressed in the same manner as the SM contributions~\cite{Buchalla:2008jp,Fajfer:2007dy}.
The most stringent limit published so far is $\BF(\Dpmmos) < 3.9 \times 10^{-6}$ (90\% \cl) by the D0 collaboration~\cite{Abazov:2007aj}.
The FOCUS collaboration places the most stringent limit on the \Ds weak annihilation decay with $\BF(\Dsmmos) < 2.6 \times 10^{-5}$~\cite{Link:2003qp}.

Lepton number violating (LNV) processes such as \Dpmmss (shown in Fig.~\ref{fig:FDs}(d)) are forbidden in the SM, because they may only occur through lepton mixing facilitated by a non-SM particle such as a Majorana neutrino~\cite{Majorana:1937vz}. 
The most stringent limits on the analysed decays at 90\% \cl are $\BF(\Dpmmss) < 2 \times 10^{-6}$ and $\BF(\Dbmmss) < 1.4 \times 10^{-5}$ set by the \babar collaboration~\cite{Lees:2011hb}.
\B meson decays set the most stringent limits on LNV decays in general, with $\BF(\Bp \to \pim \mup \mup) < 1.3 \times 10^{-8}$ at 95\% \cl set by the \lhcb collaboration~\cite{LHCB-PAPER-2011-038}.

This Letter presents the results of a search for \Dbmmos and \Dbmmss decays using $pp$ collision data, corresponding to an integrated luminosity of 1.0 \invfb, at $\sqrt{s}=7$ \tev recorded by the LHCb experiment.
The signal channels are normalised to the control channels $\Dbp\to\pip\phi$ with $\phi\to\mumu$, which have branching fraction products of \\$\BF(\Dp\to\pip(\phi\to\mumu))=(1.60\pm0.13) \times 10^{-6}$ and \\$\BF(\Ds\to\pip(\phi\to\mumu))=(1.29\pm0.14) \times 10^{-5}$~\cite{PDG2012}.

\section{The \lhcb detector and trigger}
\label{sec:detector}

The \lhcb detector~\cite{Alves:2008zz} is a single-arm forward
spectrometer covering the \mbox{pseudorapidity} range $2<\eta <5$,
designed for the study of particles containing \bquark or \cquark
quarks. The detector includes a high precision tracking system
consisting of a silicon-strip vertex detector surrounding the $pp$
interaction region, a large-area silicon-strip detector located
upstream of a dipole magnet with a bending power of about
$4{\rm\,Tm}$, and three stations of silicon-strip detectors and straw
drift tubes placed downstream. The combined tracking system has
momentum (\ptot) resolution $\Delta p/p$ that varies from 0.4\% at $5~\gevc$ to
0.6\% at $100~\gevc$, and impact parameter (IP) resolution of 20~\mum for
tracks with high transverse momentum (\pt). The IP is defined as the perpendicular distance between the path of a charged track and the primary $pp$ interaction vertex (PV) of the event.
Charged hadrons are identified
using two ring-imaging Cherenkov detectors~\cite{arXiv:1211-6759}. Photon, electron and
hadron candidates are identified by a calorimeter system consisting of
scintillating-pad and preshower detectors, an electromagnetic
calorimeter and a hadronic calorimeter. Muons are identified by a
system composed of alternating layers of iron and multiwire
proportional chambers. The trigger~\cite{Aaij:2012me} consists of a hardware stage, based
on information from the calorimeter and muon systems, followed by a
software stage that applies a full event reconstruction. 
It exploits the finite lifetime and relatively large mass of charm and beauty hadrons to distinguish heavy flavour decays from the dominant light quark processes.

The hardware trigger selects muons with \pt exceeding $1.48~\gevc$, and dimuons whose product of \pt values exceeds $(1.3~\gevc)^2$.
In the software trigger, at least one of the final state muons is required to have \ptot greater than $8~\gevc$, and an IP greater than 100~\mum. 
Alternatively, a dimuon trigger accepts candidates where both oppositely-charged muon candidates have good track quality, \pt exceeding $0.5~\gevc$, and \ptot exceeding $6~\gevc$.
In a second stage of the software trigger, two algorithms select \Dbmmos and \Dbmmss candidates. A generic \mumu trigger requires oppositely-charged muons with summed \pt greater than $1.5~\gevc$ and invariant mass, $m(\mumu)$, greater than $1~\gevcc$. A tailored trigger selects candidates with dimuon combinations of either charge and with no invariant mass requirement on the dimuon pair.

Simulated signal events are used to evaluate efficiencies and to train the selection. For the signal simulation, $pp$ collisions are generated using \pythia 6.4 \cite{Sjostrand:2006za} with a specific LHCb configuration \cite{LHCb-PROC-2010-056}. Decays of hadronic particles are described by \evtgen \cite{Lange:2001uf}. The interaction of the generated particles with the detector and its response are implemented using the \geant toolkit \cite{Allison:2006ve, *Agostinelli:2002hh} as described in Ref.~\cite{LHCb-PROC-2011-006}.
\section{Candidate selection}
\label{sec:candidateselection}

Candidate selection criteria are applied in order to maximise the significance of $\Dbmmos$ and $\Dbmmss$ signals. 
The \Dbp candidate is reconstructed from three charged tracks and is required to have a decay vertex of good quality and to have originated close to the PV by requiring that the IP $\chisq$ is less than 30.
The angle between the \Dbp candidate's momentum vector and the direction from the PV to the decay vertex, $\theta_\textrm{D}$, is required to be less than $0.8\degrees$.
The pion must have $\ptot$ exceeding 3000 \mevc, $\pt$ exceeding 500 \mevc, track fit \chisq/ndf less than 8 (where ndf is the number of degrees of freedom) and IP $\chisq$ exceeding 4. Where IP $\chisq$ is defined as the difference between the $\chisq$ of the PV reconstructed with and without the track under consideration.

A boosted decision tree (BDT) \cite{Breiman, *Roe} with the GradBoost algorithm \cite{Hocker:2007ht} distinguishes between signal-like and background-like candidates. This multivariate analysis algorithm is trained using simulated \Dpmmos signal events and a background sample taken from sidebands around the \Dbmmos peaks in an independent data sample of 36 \invpb collected in 2010. These data are not used further in the analysis. The BDT uses the following variables:
$\theta_\textrm{D}$; \chisq of both the decay vertex and flight distance of the \Dbp candidate; 
\ptot and \pt of the \Dbp candidate as well as of each of the three daughter tracks; 
IP \chisq of the \Dbp candidate and the daughter particles;
and the maximum distance of closest approach between all pairs of tracks in the candidate \Dbp decay.

Information from the rest of the event is also employed via an isolation variable, $A_{\pt}$, that considers the imbalance of \pt of nearby tracks compared to that of the \Dbp candidate
\begin{equation}
A_{\pt}=\frac{\pt(\Dbp)-(\sum\vec{\ptot})_\textrm{T}}{\pt(\Dbp)+(\sum\vec{\ptot})_\textrm{T}},
\end{equation}
where $\pt(\Dbp)$ is the \pt of the \Dbp meson and $(\sum\vec{\ptot})_\textrm{T}$ is the transverse component of the vector sum momenta of all charged particles within a cone around the candidate,
excluding the three signal tracks.
The cone is defined by a circle of radius 1.5 in the plane of pseudorapidity and azimuthal angle, measured in radians around the \Dbp candidate direction.
The signal \Dbp decay tends to be more isolated with a greater \pt asymmetry than combinatorial background.

The trained BDT is then used to classify each candidate. An optimisation study is performed to choose the combined BDT and particle identification (PID) selection criteria that maximise the expected statistical significance assuming a branching fraction of $1 \times 10^{-9}$.
The PID information is quantified as the difference in the log-likelihood under different particle mass hypotheses (DLL). The optimal cuts are found to be a BDT selection exceeding 0.9 and \dllmupi (the difference between the muon-pion hypotheses) exceeding 1 for each \mmu candidate. 

In addition, the pion candidate is required to have both \dllmupi and \dllkpi less than 0 and the two muon candidates must not share hits in the muon stations with each other or any other muon candidates. Remaining multiple candidates in an event are arbitrated by choosing the candidate with the smallest vertex \chisq (needed in $0.1\%$ of events).

Candidates from the kinematically similar \Dppp decay form an important peaking background.
A representative sample of this hadronic background is retained with a selection that is identical to that applied to the signal except for the requirement that two of the tracks have hits in the muon system.
Since the yield of this background is sizeable, a 1\% prescale is applied.
The remaining \Dppp candidates are reconstructed under the \Dbmmos and \Dbmmss hypotheses and define the probability density function (PDF) of this peaking background in the fit to the signal samples.


\section{Invariant mass fit}
\label{sec:fit}

The shapes and yields of the signal and background contributions are determined using a binned maximum likelihood fit to the invariant mass distributions of the \Dbmmos and \Dbmmss candidates in the range $1810-2040$~\mevcc.
This range is chosen to fully contain the PDFs of the correctly identified \Dp and \Ds candidates as well as those of \Dppp decays misidentified as \Dbmmos or \Dbmmss.

The \Dbmmos and \Dbmmss signals are described by the function
\begin{equation}
f(x) \propto \exp\left(\frac{-(x-\mu)^2}{2\sigma^2+(x-\mu)^2\alpha_{L,R}}\right),
\label{eq:cruijff}
\end{equation}
which is a Gaussian-like peak of mean $\mu$, width $\sigma$ and where $\alpha_L(x<\mu)$ and $\alpha_R(x>\mu)$ parameterise the tails.
The parameters of this shape are determined simultaneously across all bins (discussed below) of a given fit including the bin containing the control mode.

The \Dppp peaking background data are also split into the predefined regions and fitted with Eq.~\ref{eq:cruijff}. This provides a high-statistics, well-defined shape for this prominent background, which is imported into the corresponding subsample signal fit. The misidentification rate (the ratio of the yield in the signal data sample to that in the $\pip\pip\pim$ sample) is allowed to vary but is assumed to be constant across all bins in the fit. A systematic uncertainty is assigned to account for this assumption.
 
A second-order polynomial function is used to describe the PDF of all other combinatorial or partially reconstructed backgrounds that vary smoothly across the fit range. The coefficients of the polynomial are permitted to vary independently in each bin.

The \Dbmmos and \Dbmmss data are split into bins of \Mmumu and \Mpimu, respectively. The bins are chosen such that the resonances present in \Mmumu in the case of \Dbmmos are separate from the regions sensitive to the signal, which are in the ranges $250-525 \, \mevcc$ and $1250-2000 \, \mevcc$.
For the \Dbmmss search, the bins of \Mpimu improve the statistical significance of any signal observed, as it is assumed that a Majorana neutrino would only appear in one subsample.
The definitions of these subsamples are provided in Tables~\ref{tab:yields1} and~\ref{tab:yields2}. Cross-feed between the bins is found to be negligible from simulation studies.

The \Dbmmos and \Dbmmss data are fitted independently, with the \Dbmmos sample being fitted in two parts due to the requirement of some of the software triggers that $m(\mumu)$ exceeds $1.0~\gevcc$.
A \Dbmmos fit excluding these trigger lines simultaneously fits the low-\Mmumu, \Peta, \Prho/\Pomega and \Pphi bins.
Another fit to the \Dbmmos data, including these trigger lines, is applied to the high-\Mmumu and \Pphi bins. The \Pphi bin is needed as it provides a signal shape and normalises any signal yield. 
A simultaneous fit to the \Dbmmss data is done in all four \Mpimu bins. The \Pphi bin from the \Dbmmos data is again used to provide a signal shape and to normalise any signal yield.

\begin{figure*}[htp]
\centering
\labellist \small\hair 2pt \pinlabel (a) at 130 315 \endlabellist
\includegraphics[width=0.48\textwidth]{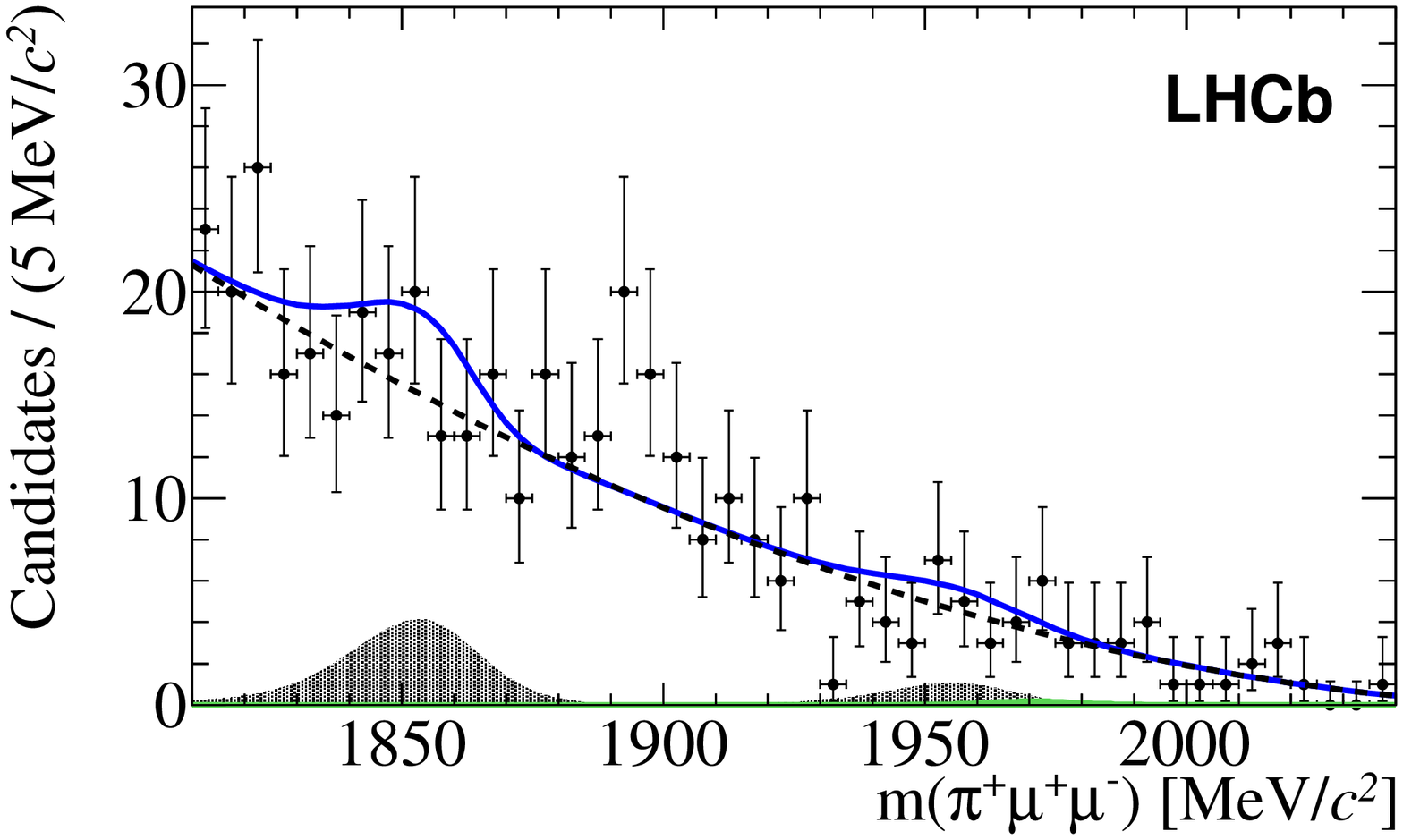}
\\
\labellist \small\hair 2pt \pinlabel (b) at 130 315 \endlabellist
\includegraphics[width=0.48\textwidth]{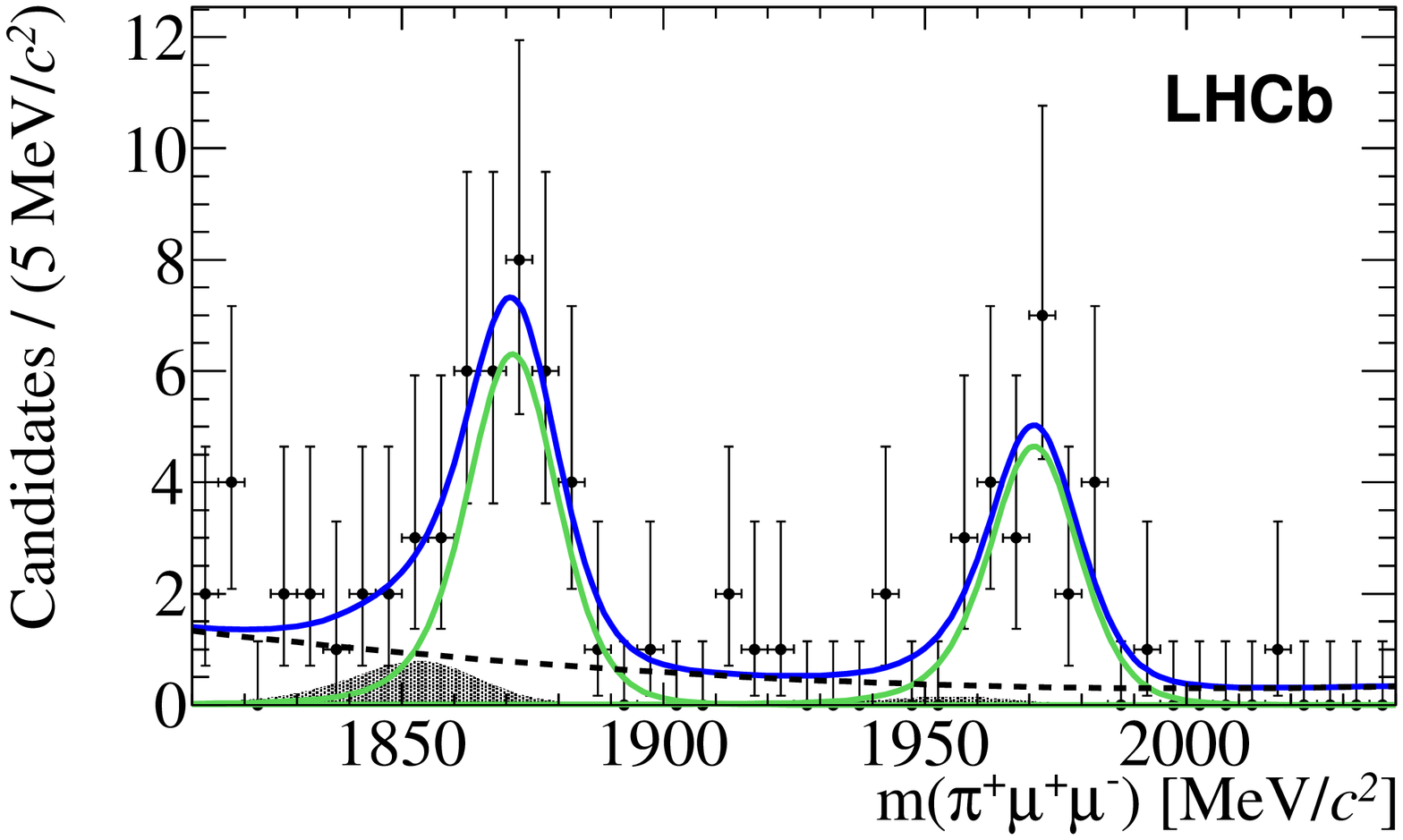}
\hfill
\labellist \small\hair 2pt \pinlabel (c) at 130 315 \endlabellist
\includegraphics[width=0.48\textwidth]{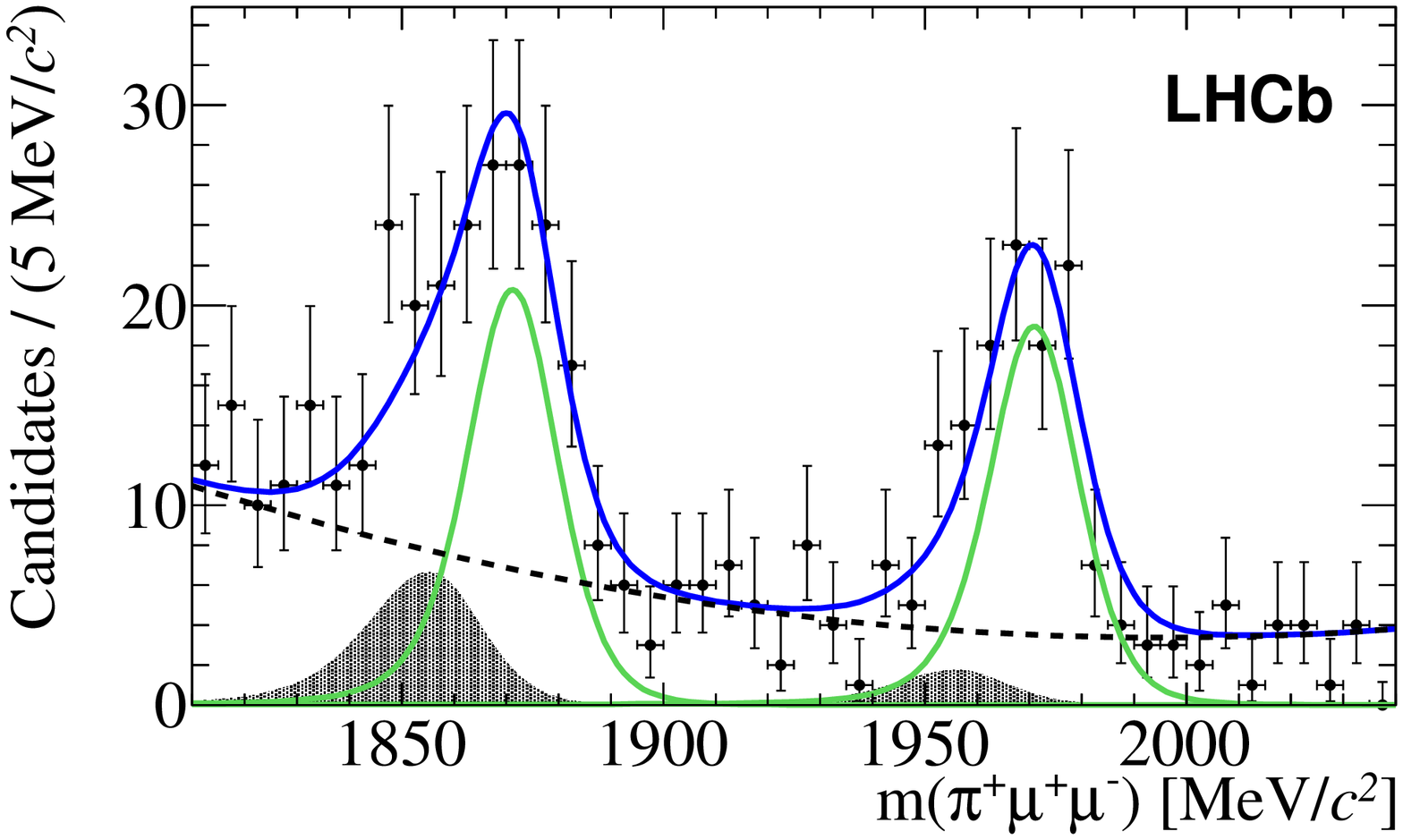}
\hfill
\labellist \small\hair 2pt \pinlabel (d) at 130 315 \endlabellist
\includegraphics[width=0.48\textwidth]{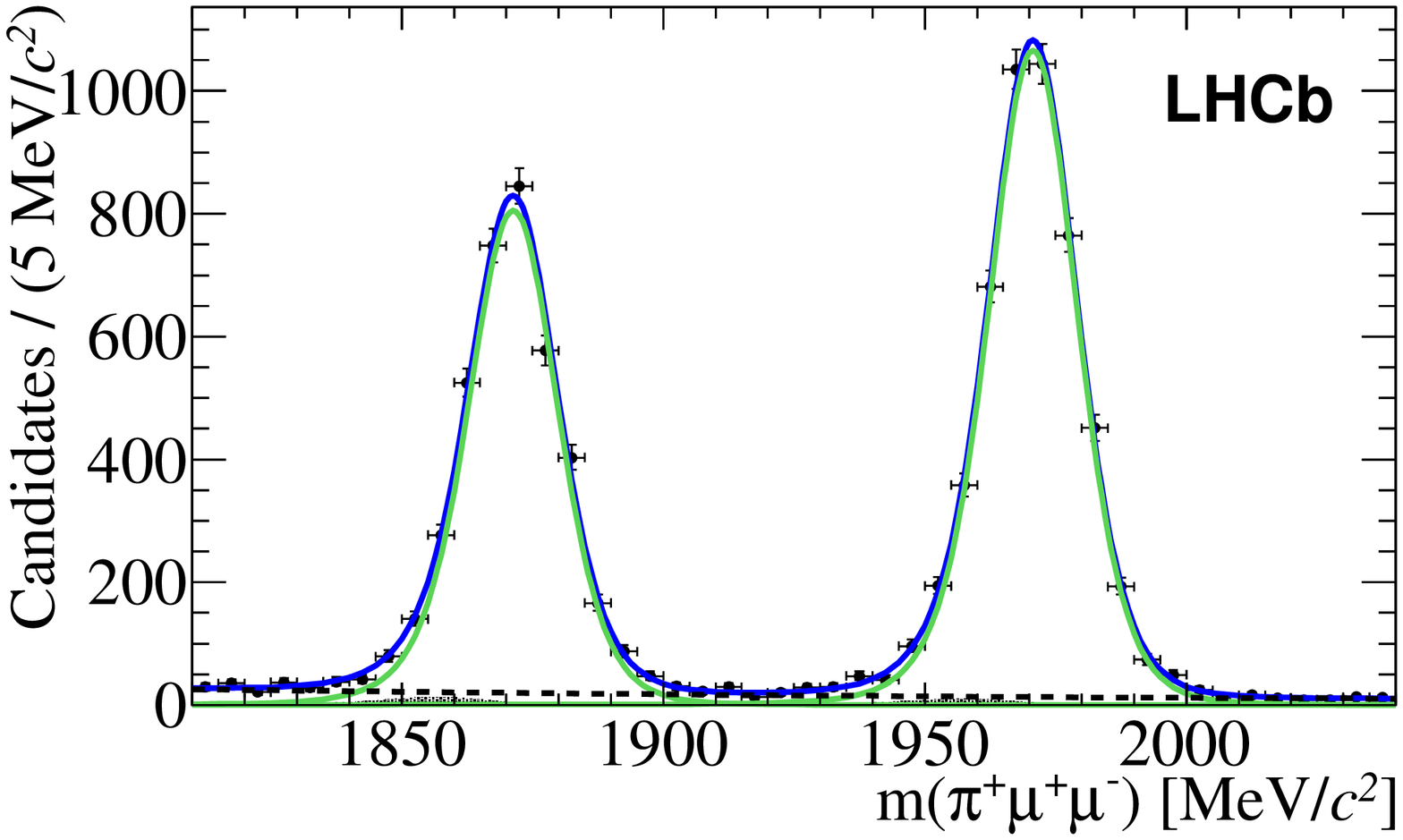}
\hfill
\labellist \small\hair 2pt \pinlabel (e) at 130 315 \endlabellist
\includegraphics[width=0.48\textwidth]{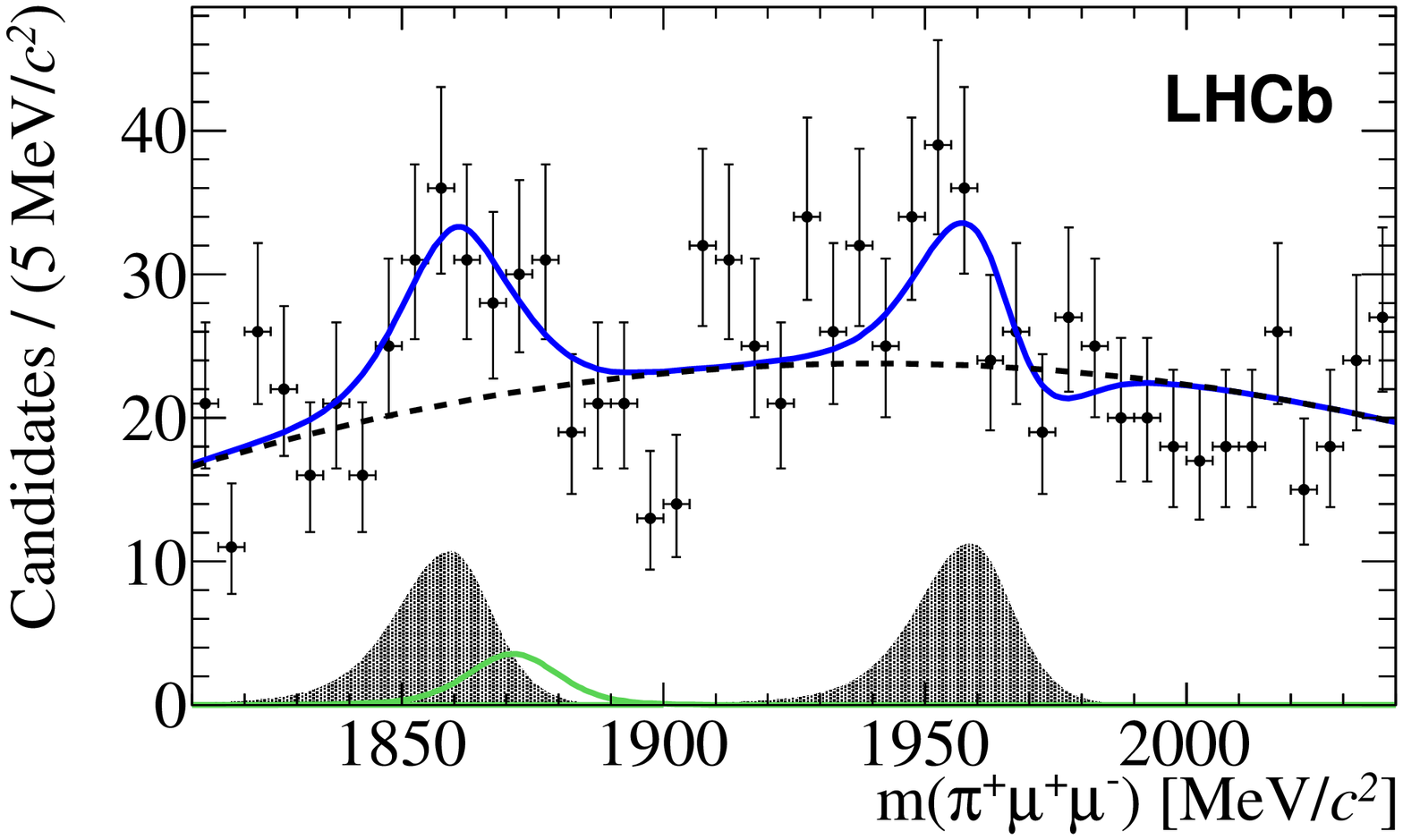}
\caption{\small{Invariant mass distributions for \Dbmmos candidates in the five $m(\mumu)$ bins. Shown are the (a) low-\Mmumu, (b) \Peta, (c) \Prho/\Pomega, (d) \Pphi (including trigger lines with $\Mmumu>1.0~\gevcc$), and (e) high-\Mmumu regions. The data are shown as points (black) and the total PDF (dark blue line) is overlaid. The components of the fit are also shown: the signal (light green line), the peaking background (solid area) and the non-peaking background (dashed line).}}
\label{fig:mass1}
\end{figure*}

\begin{figure}[htp]
\centering
\labellist \small\hair 2pt \pinlabel (a) at 130 315 \endlabellist
\includegraphics[width=0.48\textwidth]{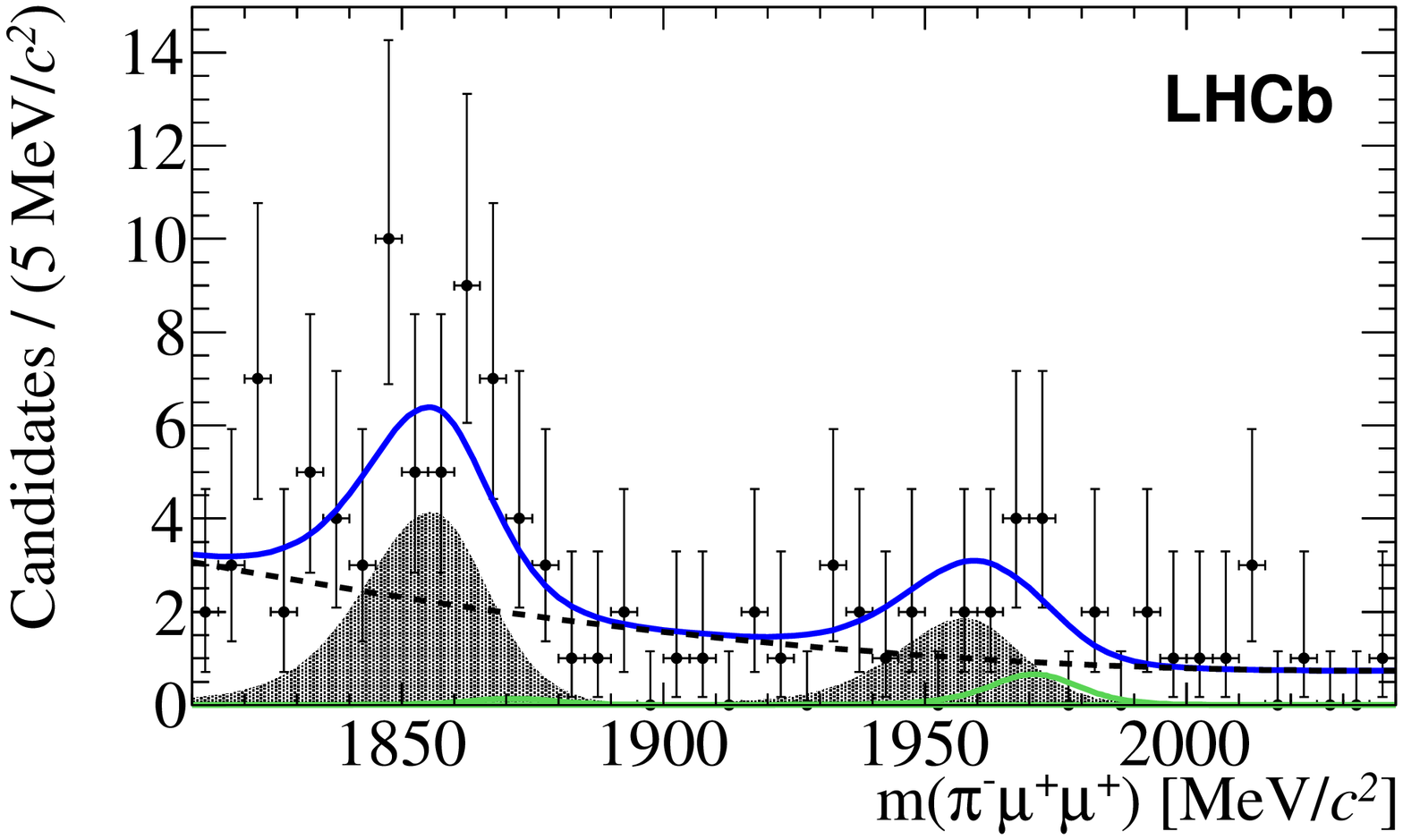}
\hfill
\labellist \small\hair 2pt \pinlabel (b) at 130 315 \endlabellist
\includegraphics[width=0.48\textwidth]{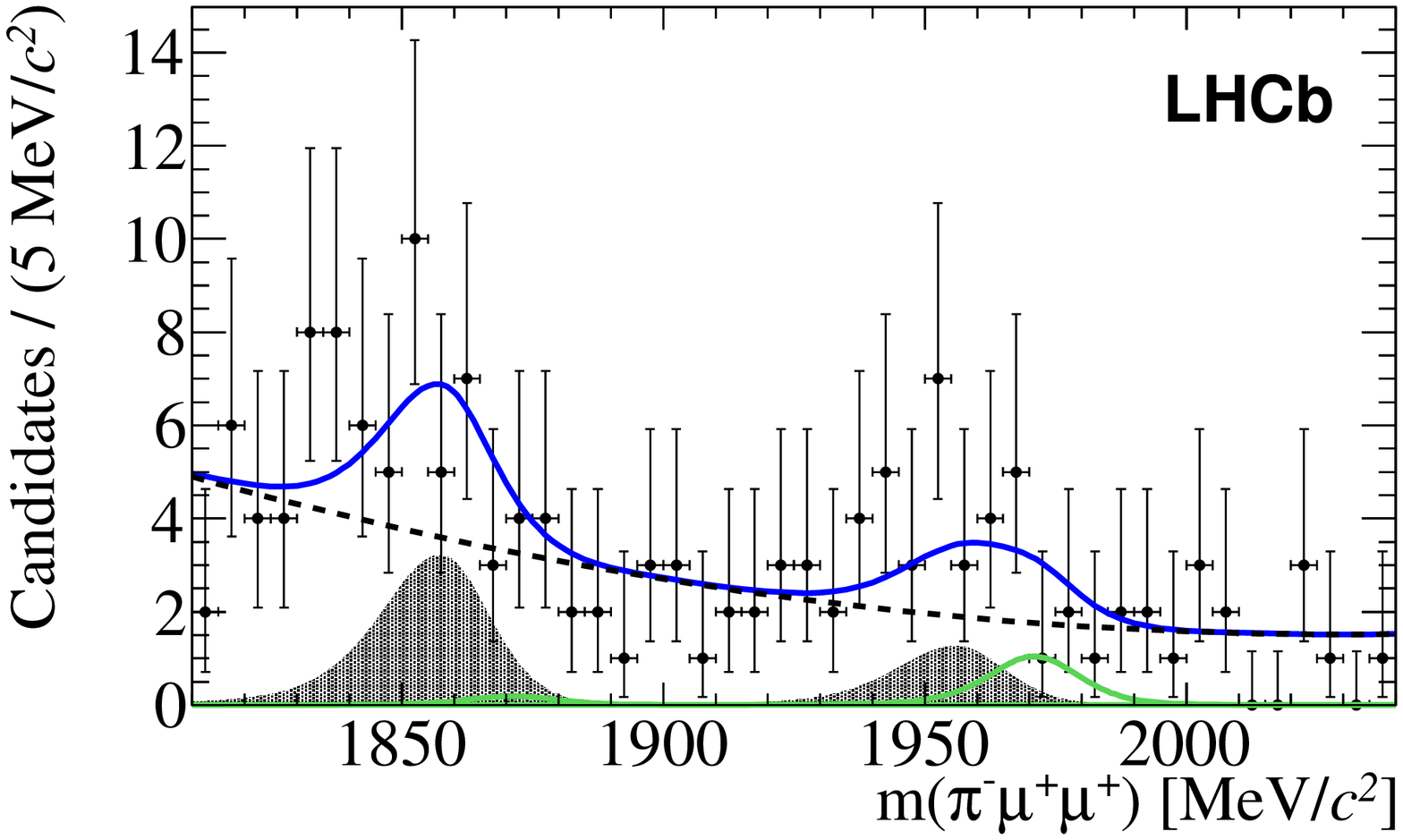}
\hfill
\labellist \small\hair 2pt \pinlabel (c) at 130 315 \endlabellist
\includegraphics[width=0.48\textwidth]{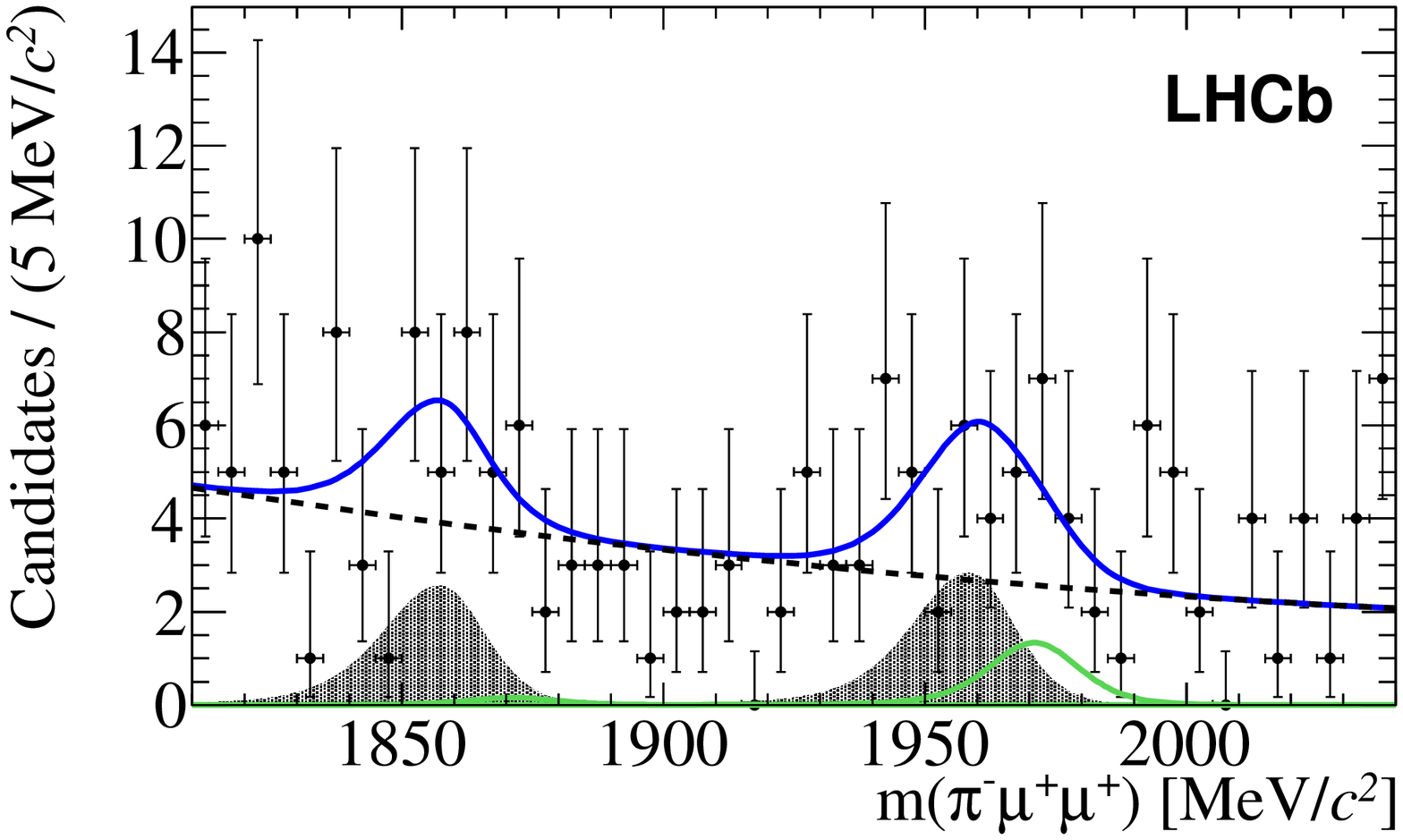}
\hfill
\labellist \small\hair 2pt \pinlabel (d) at 130 315 \endlabellist
\includegraphics[width=0.48\textwidth]{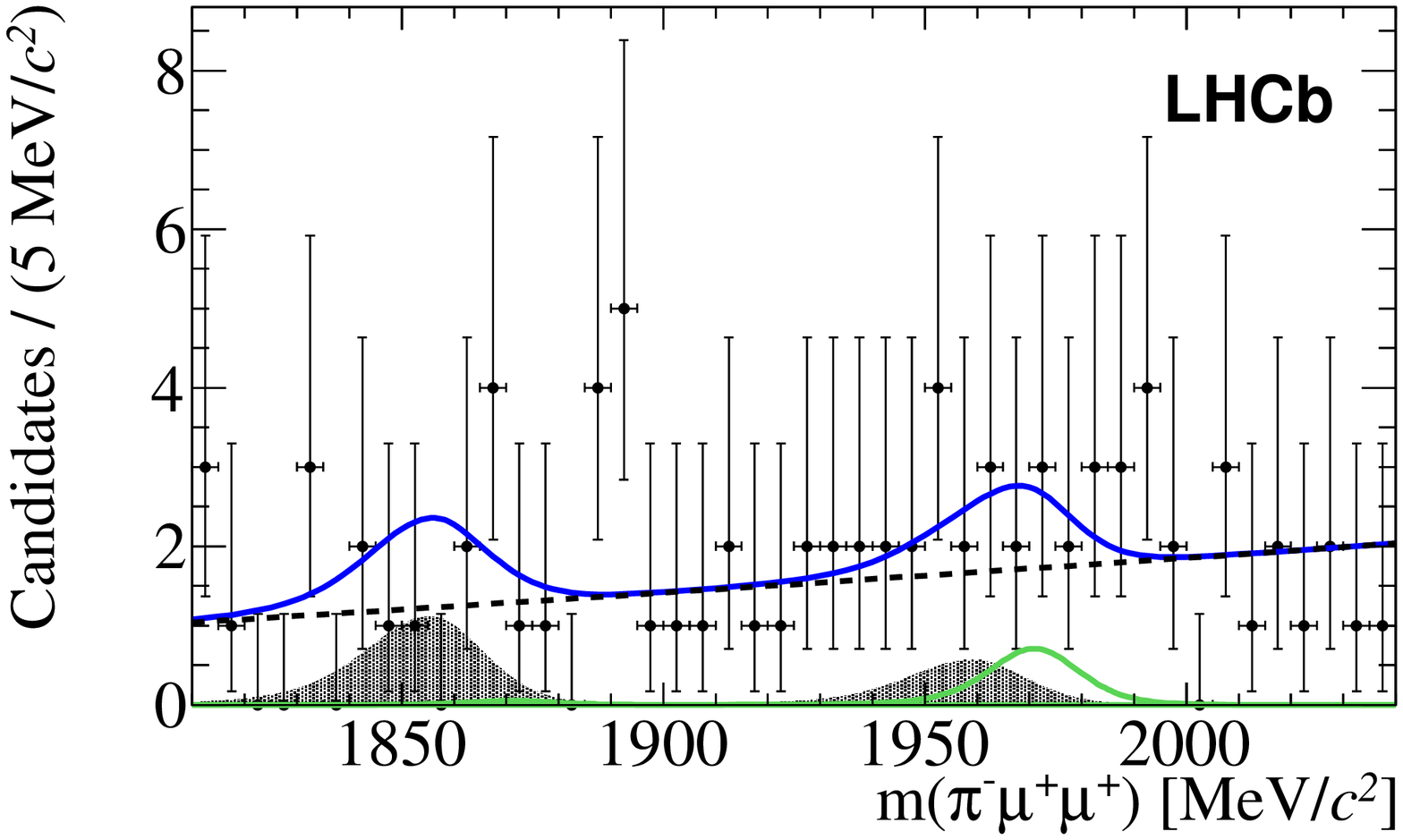}
\caption{\small{Invariant mass distributions for \Dbmmss in the four $m(\pim\mup)$ regions. Shown are (a) bin 1, (b) bin 2, (c) bin 3, and (d) bin 4. The data are shown as black points and the total PDF (dark blue line) is overlaid. The components of the fit are also shown: the signal (light green line), the peaking background (solid area) and the non-peaking background (dashed line).}}
\label{fig:mass2}
\end{figure}

\begin{figure*}[hbp]
\centering
\labellist \small\hair 2pt \pinlabel (a) at 130 315 \endlabellist
\includegraphics[width=0.48\textwidth]{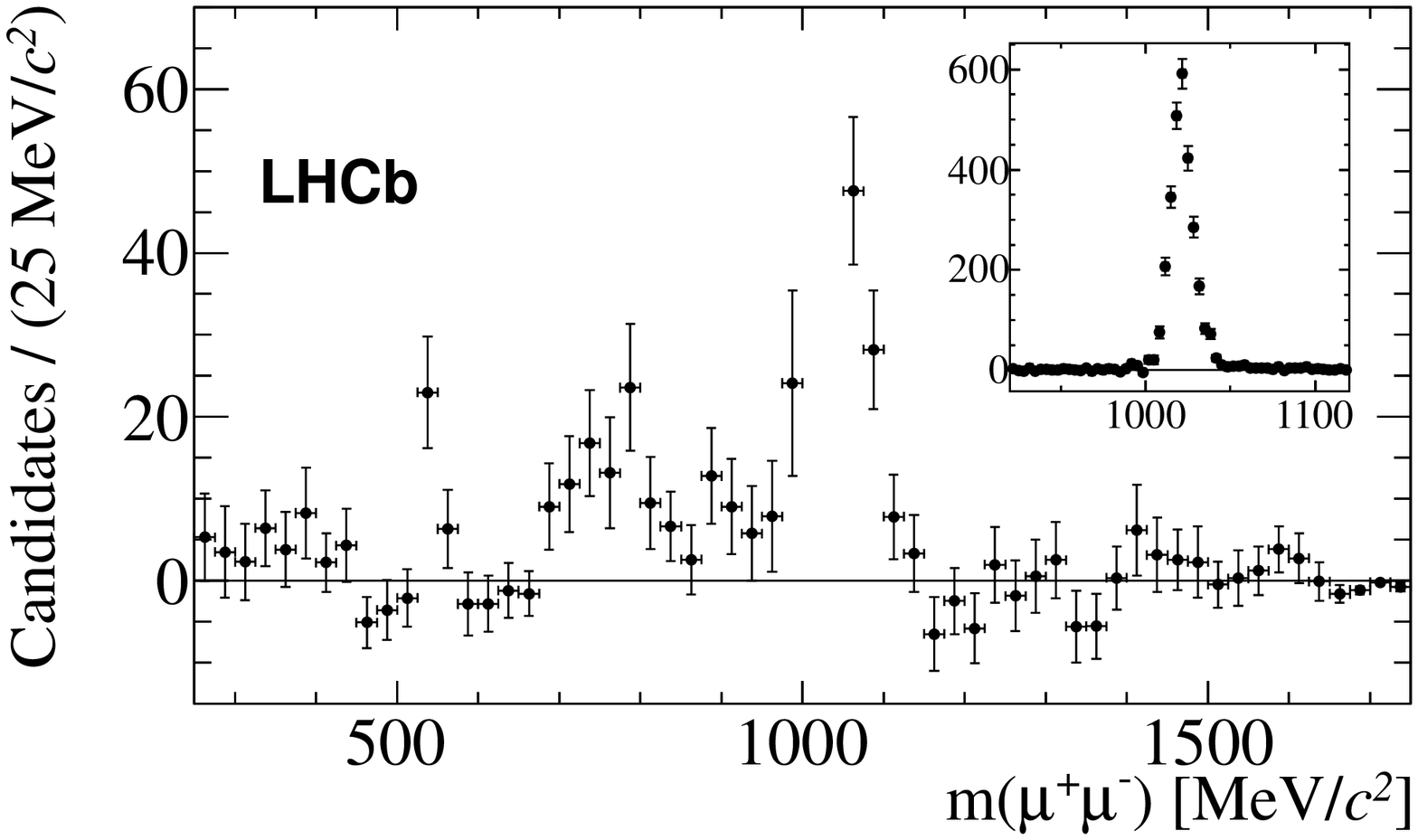}
\hfill
\labellist \small\hair 2pt \pinlabel (b) at 130 315 \endlabellist
\includegraphics[width=0.48\textwidth]{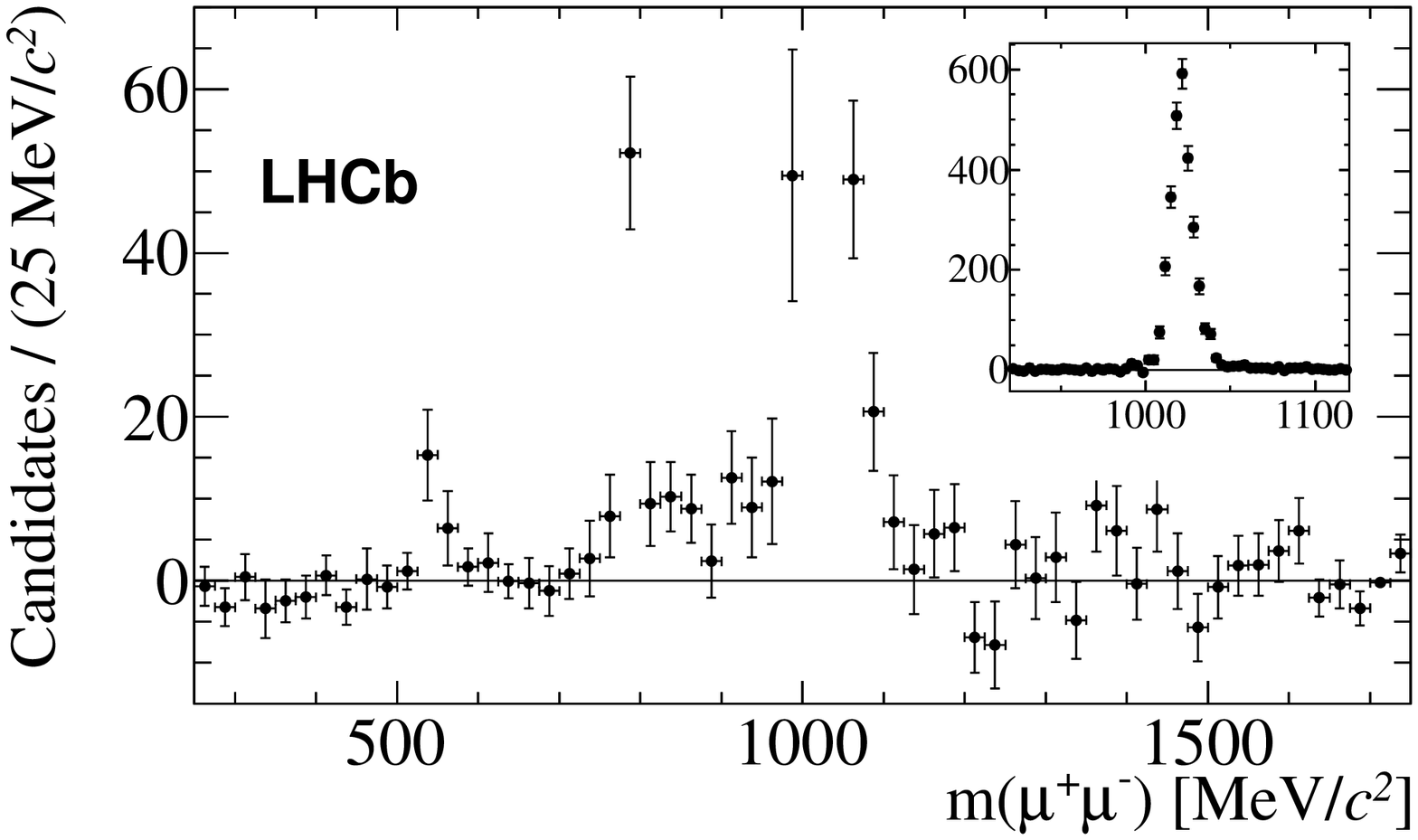}\\
\caption{\small{Background-subtracted m(\mumu) spectrum of (a) \Dpmmos and (b) \Dsmmos candidates that pass the final selection. The inset shows the \Pphi contribution, and the main figure shows the \Peta and the \Prho/\Pomega contributions. The non-peaking structure of the low and high-\Mmumu regions is also visible.}}
\label{fig:mMuMu}
\end{figure*}

The invariant mass spectra together with the results are shown in Figs.~\ref{fig:mass1} and \ref{fig:mass2}.
Background-subtracted \Mmumu distributions are obtained using the $sPlot$ technique~\cite{Pivk:2004ty} and shown in Fig.~\ref{fig:mMuMu}.
The signal yields are shown in Table~\ref{tab:yields1} for \Dbmmos decays, and in Table~\ref{tab:yields2} for \Dbmmss decays. The statistical significances of the two observed peaks are found by performing the fit again with the background-only hypothesis. Significances of 6.1 and 6.2 $\sigma$ are found for \Dpetapi and \Dsetapi decays, respectively. In comparison to \Dphipi, $\BF(\Dpetapi)=(2.2\pm0.6)\times10^{-8}$ and $\BF(\Dsetapi)=(6.8\pm2.1)\times10^{-8}$ for the \Dp and \Ds decays, respectively, and match those expected based on the $\Dbp\to\Peta\pip$ and $\Peta\to\mumu$ branching fractions \cite{PDG2012}. No significant excess of candidates is seen in any of the signal search windows.

\begin{table}[htp]
\footnotesize
\centering
\caption{\small{Signal yields for the \Dbmmos fits. The \Pphi region yields differ due to the different trigger conditions.}}
\begin{tabular}{ccccc}
Trigger conditions & Bin description & \Mmumu range [$\mevcc$] & \Dp yield & \Ds yield\\
\hline
& low-\Mmumu & $\phantom{1}250-\phantom{1}525$ & $\phantom{00}-3\pm11$ & $\phantom{-000}1\pm\phantom{0}6$\\
Triggers without & \Peta & $\phantom{1}525-\phantom{1}565$ & $\phantom{-00}29\pm\phantom{0}7$ & $\phantom{-00}22\pm\phantom{0}5$\\
$\Mmumu>1.0~\gevcc$ & \Prho/\Pomega & $\phantom{1}565-\phantom{1}850$ & $\phantom{-00}96\pm15$ & $\phantom{-00}87\pm12$\\
& \Pphi & $\phantom{1}850-1250$ & $\phantom{-}2745\pm67$ & $\phantom{-}3855\pm86$\\
\hline
\multirow{2}{*}{All triggers} & \Pphi & $\phantom{1}850-1250$ & $\phantom{-}3683\pm90$ & $\phantom{-}4857\pm90$\\
& high-\Mmumu & $1250-2000$ & $\phantom{-00}16\pm16$ & $\phantom{0}-17\pm16$\\
\end{tabular}
\label{tab:yields1}
\end{table}


\begin{table}[htp]
\footnotesize
\centering
\caption{\small{Signal yields for the \Dbmmss fit. The \Pphi region from the \Dbmmos channel is used for normalisation. The particle '$x$' is a $\Ppi$ when referring to \Dbmmss data and a $\Pmu$ for \Dbmmos data.}}
\begin{tabular}{cccc}
Bin description & $m(\mup x^-)$ range [$\mevcc$] & \Dp yield & \Ds yield\\
\hline
\Pphi & $\phantom{1}850-1250$ & $\phantom{-}2771\pm65$ & $\phantom{-}3885\pm85$\\
bin 1 & $\phantom{1}250-1140$ & $\phantom{-000}7\pm\phantom{0}6$ & $\phantom{-000}4\pm\phantom{0}4$\\
bin 2 & $1140-1340$ & $\phantom{00}-3\pm\phantom{0}6$ & $\phantom{-000}3\pm\phantom{0}5$\\
bin 3 & $1340-1550$ & $\phantom{00}-1\pm\phantom{0}6$ & $\phantom{-000}6\pm\phantom{0}6$\\
bin 4 & $1540-2000$ & $\phantom{-000}0\pm\phantom{0}4$ & $\phantom{-000}4\pm\phantom{0}5$\\
\end{tabular}
\label{tab:yields2}
\end{table}


\section{Branching fraction determination}
\label{sec:branchingfractiondetermination}

The \Dbmmos and \Dbmmss branching fractions are calculated using
\begin{equation}
{\BF(\Dpmmns)}
= \frac{N_{\Dpmmns}}{N_{\Dphipi}}
\times \frac{\epsilon_{\Dphipi}}{\epsilon_{\Dpmmns}}
\times {\BF(\Dphipi)},
\label{Eq:ExtractBR}
\end{equation}
where \Dpmmns represents either \Dbmmos or \Dbmmss.
The relevant signal yield and efficiency are given by $N_{\Dpmmns}$ and $\epsilon_{\Dpmmns}$, respectively, and the relevant control mode yield and efficiency are given by $N_{\Dphipi}$ and $\epsilon_{\Dphipi}$, respectively.

The efficiency of the signal decay mode and the control mode include the efficiencies of the geometrical acceptance of the detector, track reconstruction, muon identification, selection, and trigger.
The accuracy with which the simulation reproduces the track reconstruction and identification is limited.
For that reason, the corresponding efficiencies are also studied in real data.
A tag and probe technique applied to $ B \rightarrow J/\psi X$ decays provides a large sample of unambiguous muons to determine the tracking and muon identification efficiencies.
The pion identification is studied using $\Dstarp \rightarrow  \pip (\Dz \rightarrow \Km \pip)$  decays.
The efficiencies observed as a function of the particle momentum and pseudorapidity and of the track multiplicity in the event are used to correct the efficiencies determined by the simulation.
The correction to the efficiency ratio is typically of the order of 2\% in each \Mmumu or \Mpimu region.
Small relative corrections are expected since the signal and control modes share almost identical final states.


\clearpage
\section{Systematic uncertainties}
\label{sec:systematics}

Systematic uncertainties in the calculation of the signal branching fractions arise due to imperfect knowledge of the control mode branching fraction, the efficiency ratio, and the yield ratio.
 
A systematic uncertainty of the order 10\% accompanies the branching fraction of the control mode \Dphipi and is the dominant source of the systematic uncertainty on the branching fraction measurement.

A systematic uncertainty affecting the efficiency ratio is due to
the geometrical acceptance of the detector, which depends on the angular distributions of the final state particles, and thus on the decay model.
By default, signal decays are simulated with a phase-space model.
A conservative 1\% uncertainty is determined by recalculating the acceptance assuming a flat \Mmumu distribution.

The uncertainties on the tracking and particle identification corrections also affect the efficiency ratio and involve statistical components due to the size of the data samples and systematic uncertainties inherent in the techniques employed to determine the corrections.
The corrections depend upon the choice of control sample, the selection and trigger requirements applied to this sample, and the precise definition of the probe tracks.
The binning used to weight the efficiency as a function of the momentum, pseudorapity and multiplicity is varied to evaluate the uncertainty.
The uncertainty in the choice of phase space model is accounted for by comparing the efficiency corrections in the extreme bins of the \Mmumu or \Mpimu distributions.
In total, the uncertainty due to particle reconstruction and identification is found to be 4.2\% across all bins.

Also affecting the efficiency ratio is the fact that the offline selection is not perfectly described by simulation.
The systematic uncertainty is estimated by smearing track properties to reproduce the distributions observed in data, using \Dphipi decays as a reference.
The corresponding variation in the efficiency ratio indicates an uncertainty of 4\%. 
Also, the trigger requirements imposed to select the signal are varied in order to test the imperfect simulation of the online reconstruction and $3\%$ uncertainty is deduced.
The sources of uncertainty discussed so far are given in Table~\ref{tab:systematics1}.

Final uncertainty on the efficiency ratio arises due to the finite size of the simulated samples. It is calculated separately in each \Mmumu and \Mpimu bin.
These contributions are included in the systematic uncertainties shown in Table~\ref{tab:systematics2}.


The systematic uncertainties affecting the yield ratio are taken into account when the branching fraction limits are calculated.
The shapes of the signal peaks are assumed to be the same in all \Mmumu and \Mpimu bins.
A 10\% variation of the width of the Gaussian-like PDF, seen in simulation, is taken into account for variation across the bins.
In each bin, the shape of the \Dppp peaking background is taken from a simultaneous fit to a larger sample to which looser \dllmupi criteria is applied. As simulation shows the shape of the PDF is altered by a \dllmupi requirement.
A variation in the peaking background's fitted width equal to 20\% is applied as a systematic uncertainty.
The pion-to-muon misidentification rate is assumed to be the same in all bins. Simulation suggests that a systematic variation of 20\% in this quantity is conservative.
Contributions to the yield ratio systematic uncertainty are found to increase the upper limit on the branching fraction by around 10\%.

\begin{table}[ht]
\footnotesize
\centering 
\caption{\small{Relative systematic uncertainties averaged over all bins and decay modes for the control mode branching fraction and efficiency ratio. The number in parentheses refers to the \Ds decay.}} 
\begin{tabular}{lc} 
Source &  Uncertainty (\%)\\
\hline 
Geometric acceptance & 1.0 \\
Track reconstruction and PID & 4.2 \\
Stripping and BDT efficiency & 4.0 \\
Trigger efficiency & 3.0 \\
\BF(\Dbp\to\pip\Pphi(\mumu)) uncertainty & 8.1 (10.9) \\
\end{tabular}
\label{tab:systematics1} 
\end{table}

\begin{table}[ht]
\footnotesize
\centering 
\caption{\small{Total systematic uncertainty in each \Mmumu and \Mpimu bin with the uncertainty on the control mode branching fraction, the efficiency ratio and the statistical uncertainty stemming from the size of the simulated samples  added in quadrature. The numbers in parentheses refer to the \Ds decay.}}
\begin{tabular}{ccc} 
Bin description &  \Dbmmos (\%) & \Dbmmss (\%) \\
\hline 
low-\Mmumu  &  11.8 (16.9) & \\ 
high-\Mmumu  & 11.2 (15.5)  \\ 	
bin 1             & &  11.1 (17.0) \\ 
bin 2            &  & 10.9 (16.4) \\ 
bin 3            &  & 11.1 (16.0) \\ 
bin 4           &  & 11.3 (16.0) \\  [1ex] 
\end{tabular}
\label{tab:systematics2} 
\end{table}
\section{Results}
\label{sec:results}


The compatibility of the observed distribution of candidates with a signal plus background or background-only hypothesis is evaluated using the \cls method~\cite{CLsMethod, Junk:1999kv}.
The method provides two estimators: \cls, a measure of the compatibility of the observed distribution with the signal hypothesis, and \clb, a measure of the compatibility with the background-only hypothesis. 
The systematic uncertainties are included in the \cls method using the techniques described in Ref.~\cite{CLsMethod, Junk:1999kv}.

\begin{figure}[htp]
\centering
\includegraphics[scale=0.4]{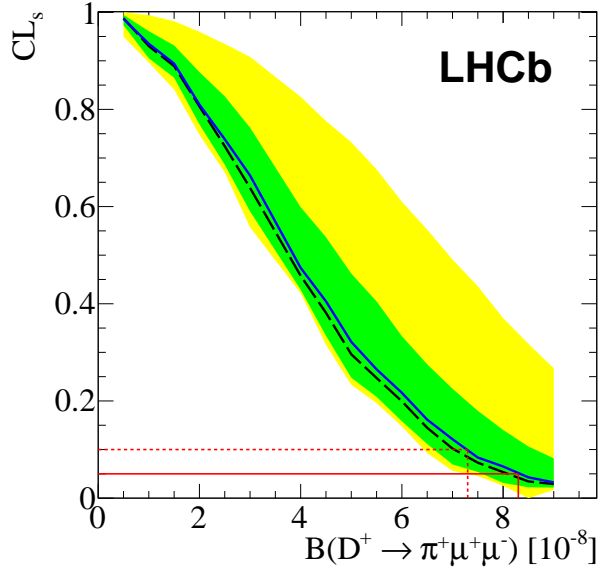}
\caption{\small{Observed (solid curve) and expected (dashed curve) \cls values as a function of \BF(\Dpmmos). The green (yellow) shaded area contains the $\pm1\sigma$ ($\pm2\sigma$) interval of possible results compatible with the expected value if only background is observed. The upper limits at the 90\% (95\%) \cl are indicated by the dashed (solid) line.}}
\label{fig:cls}
\end{figure}

Upper limits on the \Dpmmos and \Dpmmss branching fractions are determined using the observed distribution of \cls as a function of the branching fraction in each \Mmumu or \Mpimu bin.
Total branching fractions are found using the same method and by considering the fraction of simulated signal candidates in each \Mmumu or \Mpimu bin.
The simulated signal assumes a phase-space model for the non-resonant decays.
The observed distribution of \cls as a function of the total branching fraction for \Dpmmos is shown in Fig.~\ref{fig:cls}.
The upper limits at 90\% and 95\% \cl and the p-values $(1-\clb)$ for the the background-only hypothesis are shown in Table~\ref{tab:limits1}	.


\begin{table*}
\footnotesize
\centering
\caption{\small{Upper limits in each \Mmumu and \Mpimu bin and total branching fractions at the 90\% and 95\% \cl and p-values for the background-only hypothesis.}}
\begin{tabular}{ccccc}
Decay & Bin & $90\% \, [\times10^{-8}]$ & $95\% \, [\times10^{-8}]$ & p-value\\
\hline
\multirow{3}{*}{\Dpmmos} & low-\Mmumu & 2.0 & 2.5 & 0.74\\
& high-\Mmumu & 2.6 & 2.9 & 0.42\\
& Total & 7.3 & 8.3 & 0.42\\
\hline
\multirow{3}{*}{\Dsmmos} & low-\Mmumu & 6.9 & 7.7 & 0.78\\
& high-\Mmumu & 16.0 & 18.6 & 0.41\\
& Total & 41.0 & 47.7 & 0.42\\
\hline
\multirow{5}{*}{\Dpmmss} & bin 1 & 1.4 & 1.7 & 0.32\\
& bin 2 & 1.1 & 1.3 & 0.61\\
& bin 3 & 1.3 & 1.5 & 0.94\\
& bin 4 & 1.3 & 1.5 & 0.97\\
& Total & 2.2 & 2.5 & 0.86\\
\hline
\multirow{5}{*}{\Dsmmss} & bin 1 & 6.2 & 7.6 & 0.34\\
& bin 2 & 4.4 & 5.3 & 0.51\\
& bin 3 & 6.0 & 7.3 & 0.32\\
& bin 4 & 7.5 & 8.7 & 0.41\\
& Total & 12.0 & 14.1 & 0.12\\
\end{tabular}
\label{tab:limits1}
\end{table*}

\clearpage
\section{Conclusions}
\label{sec:conclusions}

A search for the \Dbmmos and \Dbmmss decays has been conducted using proton-proton collision data, corresponding to an integrated luminosity of 1.0 \invfb, at $\sqrt{s}=7$ \tev recorded by the LHCb experiment. Limits are set on branching fractions in several $m(\mumu)$ and $m(\pim\mup)$ bins and on the total branching fraction excluding the resonant contributions assuming a phase-space model. These results are the most stringent to date and represent an improvement by a factor of fifty compared to previous results. The observed data, away from resonant structures, is compatible with the background-only hypothesis, and no enhancement is observed. The $90\% \, (95)\%$  \cl limits on the branching fractions are

\begin{eqnarray*}
\mathcal{\BF}(\Dpmmos) < 7.3 \, (8.3) \times 10^{-8},\\
\mathcal{\BF}(\Dsmmos) < 4.1 \, (4.8) \times 10^{-7},\\
\mathcal{\BF}(\Dpmmss) < 2.2 \, (2.5) \times 10^{-8},\\
\mathcal{\BF}(\Dsmmss) < 1.2 \, (1.4) \times 10^{-7}.
\end{eqnarray*}
\clearpage
\section*{Acknowledgements}

\noindent We would like to thank Nejc Ko\v{s}nik for very useful discussions on the
theoretical aspects of the decay modes studied in this paper.
We express our gratitude to our colleagues in the CERN
accelerator departments for the excellent performance of the LHC. We
thank the technical and administrative staff at the LHCb
institutes. We acknowledge support from CERN and from the national
agencies: CAPES, CNPq, FAPERJ and FINEP (Brazil); NSFC (China);
CNRS/IN2P3 and Region Auvergne (France); BMBF, DFG, HGF and MPG
(Germany); SFI (Ireland); INFN (Italy); FOM and NWO (The Netherlands);
SCSR (Poland); ANCS/IFA (Romania); MinES, Rosatom, RFBR and NRC
``Kurchatov Institute'' (Russia); MinECo, XuntaGal and GENCAT (Spain);
SNSF and SER (Switzerland); NAS Ukraine (Ukraine); STFC (United
Kingdom); NSF (USA). We also acknowledge the support received from the
ERC under FP7. The Tier1 computing centres are supported by IN2P3
(France), KIT and BMBF (Germany), INFN (Italy), NWO and SURF (The
Netherlands), PIC (Spain), GridPP (United Kingdom). We are thankful
for the computing resources put at our disposal by Yandex LLC
(Russia), as well as to the communities behind the multiple open
source software packages that we depend on.

\addcontentsline{toc}{section}{References}
\bibliographystyle{LHCb}
\bibliography{main}

\ifx\mcitethebibliography\mciteundefinedmacro
\PackageError{LHCb.bst}{mciteplus.sty has not been loaded}
{This bibstyle requires the use of the mciteplus package.}\fi
\providecommand{\href}[2]{#2}
\begin{mcitethebibliography}{10}
\mciteSetBstSublistMode{n}
\mciteSetBstMaxWidthForm{subitem}{\alph{mcitesubitemcount})}
\mciteSetBstSublistLabelBeginEnd{\mcitemaxwidthsubitemform\space}
{\relax}{\relax}

\bibitem{Fajfer:2006yc}
S.~Fajfer and S.~Prelovsek, \ifthenelse{\boolean{articletitles}}{{\it {Search
  for new physics in rare D decays}}, }{}Conf.\ Proc.\  {\bf C060726} (2006)
  811, \href{http://arxiv.org/abs/hep-ph/0610032}{{\tt
  arXiv:hep-ph/0610032}}\relax
\mciteBstWouldAddEndPuncttrue
\mciteSetBstMidEndSepPunct{\mcitedefaultmidpunct}
{\mcitedefaultendpunct}{\mcitedefaultseppunct}\relax
\EndOfBibitem
\bibitem{Abe:2001dh}
Belle collaboration, K.~Abe {\em et~al.},
  \ifthenelse{\boolean{articletitles}}{{\it {Observation of the decay $B \to K
  \ell^{+} \ell^{-}$}},
  }{}\href{http://dx.doi.org/10.1103/PhysRevLett.88.021801}{Phys.\ Rev.\ Lett.\
   {\bf 88} (2002) 021801}, \href{http://arxiv.org/abs/hep-ex/0109026}{{\tt
  arXiv:hep-ex/0109026}}\relax
\mciteBstWouldAddEndPuncttrue
\mciteSetBstMidEndSepPunct{\mcitedefaultmidpunct}
{\mcitedefaultendpunct}{\mcitedefaultseppunct}\relax
\EndOfBibitem
\bibitem{Park:2001cv}
HyperCP collaboration, H.~Park {\em et~al.},
  \ifthenelse{\boolean{articletitles}}{{\it {Observation of the decay $\Km \to
  \pim \mumu $and measurements of the branching ratios for $\Kp \to \pip
  \mumu$}}, }{}\href{http://dx.doi.org/10.1103/PhysRevLett.88.111801}{Phys.\
  Rev.\ Lett.\  {\bf 88} (2002) 111801},
  \href{http://arxiv.org/abs/hep-ex/0110033}{{\tt arXiv:hep-ex/0110033}}\relax
\mciteBstWouldAddEndPuncttrue
\mciteSetBstMidEndSepPunct{\mcitedefaultmidpunct}
{\mcitedefaultendpunct}{\mcitedefaultseppunct}\relax
\EndOfBibitem
\bibitem{Fajfer:2001sa}
S.~Fajfer, S.~Prelovsek, and P.~Singer,
  \ifthenelse{\boolean{articletitles}}{{\it {Rare charm meson decays $D \to P\,
  l^+ l^-$ and $c \to u\, l^+ l^-$ in SM and MSSM}},
  }{}\href{http://dx.doi.org/10.1103/PhysRevD.64.114009}{Phys.\ Rev.\  {\bf
  D64} (2001) 114009}, \href{http://arxiv.org/abs/hep-ph/0106333}{{\tt
  arXiv:hep-ph/0106333}}\relax
\mciteBstWouldAddEndPuncttrue
\mciteSetBstMidEndSepPunct{\mcitedefaultmidpunct}
{\mcitedefaultendpunct}{\mcitedefaultseppunct}\relax
\EndOfBibitem
\bibitem{Fajfer:2007dy}
S.~Fajfer, N.~Kosnik, and S.~Prelovsek,
  \ifthenelse{\boolean{articletitles}}{{\it {Updated constraints on new physics
  in rare charm decays}},
  }{}\href{http://dx.doi.org/10.1103/PhysRevD.76.074010}{Phys.\ Rev.\  {\bf
  D76} (2007) 074010}, \href{http://arxiv.org/abs/0706.1133}{{\tt
  arXiv:0706.1133}}\relax
\mciteBstWouldAddEndPuncttrue
\mciteSetBstMidEndSepPunct{\mcitedefaultmidpunct}
{\mcitedefaultendpunct}{\mcitedefaultseppunct}\relax
\EndOfBibitem
\bibitem{Paul:2011ar}
A.~Paul, I.~I. Bigi, and S.~Recksiegel,
  \ifthenelse{\boolean{articletitles}}{{\it {On $D\to X_u\, l^+ l^-$ within the
  Standard Model and frameworks like the littlest Higgs model with T parity}},
  }{}\href{http://dx.doi.org/10.1103/PhysRevD.83.114006}{Phys.\ Rev.\  {\bf
  D83} (2011) 114006}, \href{http://arxiv.org/abs/1101.6053}{{\tt
  arXiv:1101.6053}}\relax
\mciteBstWouldAddEndPuncttrue
\mciteSetBstMidEndSepPunct{\mcitedefaultmidpunct}
{\mcitedefaultendpunct}{\mcitedefaultseppunct}\relax
\EndOfBibitem
\bibitem{Buchalla:2008jp}
M.~Artuso {\em et~al.}, \ifthenelse{\boolean{articletitles}}{{\it {$B$, $D$ and
  $K$ decays}},
  }{}\href{http://dx.doi.org/10.1140/epjc/s10052-008-0716-1}{Eur.\ Phys.\ J.\
  {\bf C57} (2008) 309}, \href{http://arxiv.org/abs/0801.1833}{{\tt
  arXiv:0801.1833}}\relax
\mciteBstWouldAddEndPuncttrue
\mciteSetBstMidEndSepPunct{\mcitedefaultmidpunct}
{\mcitedefaultendpunct}{\mcitedefaultseppunct}\relax
\EndOfBibitem
\bibitem{Abazov:2007aj}
D0 collaboration, V.~Abazov {\em et~al.},
  \ifthenelse{\boolean{articletitles}}{{\it {Search for
  flavor-changing-neutral-current $D$ meson decays}},
  }{}\href{http://dx.doi.org/10.1103/PhysRevLett.100.101801}{Phys.\ Rev.\
  Lett.\  {\bf 100} (2008) 101801}, \href{http://arxiv.org/abs/0708.2094}{{\tt
  arXiv:0708.2094}}\relax
\mciteBstWouldAddEndPuncttrue
\mciteSetBstMidEndSepPunct{\mcitedefaultmidpunct}
{\mcitedefaultendpunct}{\mcitedefaultseppunct}\relax
\EndOfBibitem
\bibitem{Link:2003qp}
FOCUS Collaboration, J.~Link {\em et~al.},
  \ifthenelse{\boolean{articletitles}}{{\it {Search for rare and forbidden
  three body dimuon decays of the charmed mesons \Dp and \Ds}},
  }{}\href{http://dx.doi.org/10.1016/j.physletb.2003.07.079}{Phys.\ Lett.\
  {\bf B572} (2003) 21}, \href{http://arxiv.org/abs/hep-ex/0306049}{{\tt
  arXiv:hep-ex/0306049}}\relax
\mciteBstWouldAddEndPuncttrue
\mciteSetBstMidEndSepPunct{\mcitedefaultmidpunct}
{\mcitedefaultendpunct}{\mcitedefaultseppunct}\relax
\EndOfBibitem
\bibitem{Majorana:1937vz}
E.~Majorana, \ifthenelse{\boolean{articletitles}}{{\it {Teoria simmetrica
  dell'elettrone e del positrone}},
  }{}\href{http://dx.doi.org/10.1007/BF02961314}{Nuovo Cim.\  {\bf 14} (1937)
  171}\relax
\mciteBstWouldAddEndPuncttrue
\mciteSetBstMidEndSepPunct{\mcitedefaultmidpunct}
{\mcitedefaultendpunct}{\mcitedefaultseppunct}\relax
\EndOfBibitem
\bibitem{Lees:2011hb}
\babar collaboration, J.~Lees {\em et~al.},
  \ifthenelse{\boolean{articletitles}}{{\it {Searches for rare or forbidden
  semileptonic charm decays}},
  }{}\href{http://dx.doi.org/10.1103/PhysRevD.84.072006}{Phys.\ Rev.\  {\bf
  D84} (2011) 072006}, \href{http://arxiv.org/abs/1107.4465}{{\tt
  arXiv:1107.4465}}\relax
\mciteBstWouldAddEndPuncttrue
\mciteSetBstMidEndSepPunct{\mcitedefaultmidpunct}
{\mcitedefaultendpunct}{\mcitedefaultseppunct}\relax
\EndOfBibitem
\bibitem{LHCB-PAPER-2011-038}
LHCb collaboration, R.~Aaij {\em et~al.},
  \ifthenelse{\boolean{articletitles}}{{\it {Searches for Majorana neutrinos in
  $\Bm$ decays}}, }{}\href{http://arxiv.org/abs/1201.5600}{{\tt
  arXiv:1201.5600}}\relax
\mciteBstWouldAddEndPuncttrue
\mciteSetBstMidEndSepPunct{\mcitedefaultmidpunct}
{\mcitedefaultendpunct}{\mcitedefaultseppunct}\relax
\EndOfBibitem
\bibitem{PDG2012}
Particle Data Group, J.~Beringer {\em et~al.},
  \ifthenelse{\boolean{articletitles}}{{\it {\href{http://pdg.lbl.gov/}{Review
  of particle physics}}},
  }{}\href{http://dx.doi.org/10.1103/PhysRevD.86.010001}{Phys.\ Rev.\  {\bf
  D86} (2012) 010001}\relax
\mciteBstWouldAddEndPuncttrue
\mciteSetBstMidEndSepPunct{\mcitedefaultmidpunct}
{\mcitedefaultendpunct}{\mcitedefaultseppunct}\relax
\EndOfBibitem
\bibitem{Alves:2008zz}
LHCb collaboration, A.~A. Alves~Jr. {\em et~al.},
  \ifthenelse{\boolean{articletitles}}{{\it {The \lhcb detector at the LHC}},
  }{}\href{http://dx.doi.org/10.1088/1748-0221/3/08/S08005}{JINST {\bf 3}
  (2008) S08005}\relax
\mciteBstWouldAddEndPuncttrue
\mciteSetBstMidEndSepPunct{\mcitedefaultmidpunct}
{\mcitedefaultendpunct}{\mcitedefaultseppunct}\relax
\EndOfBibitem
\bibitem{arXiv:1211-6759}
M.~Adinolfi {\em et~al.}, \ifthenelse{\boolean{articletitles}}{{\it
  {Performance of the \lhcb RICH detector at the LHC}},
  }{}\href{http://arxiv.org/abs/1211.6759}{{\tt arXiv:1211.6759}}\relax
\mciteBstWouldAddEndPuncttrue
\mciteSetBstMidEndSepPunct{\mcitedefaultmidpunct}
{\mcitedefaultendpunct}{\mcitedefaultseppunct}\relax
\EndOfBibitem
\bibitem{Aaij:2012me}
R.~Aaij {\em et~al.}, \ifthenelse{\boolean{articletitles}}{{\it The \lhcb
  trigger and its performance}, }{}\href{http://arxiv.org/abs/1211.3055}{{\tt
  arXiv:1211.3055}}, to appear in JINST\relax
\mciteBstWouldAddEndPuncttrue
\mciteSetBstMidEndSepPunct{\mcitedefaultmidpunct}
{\mcitedefaultendpunct}{\mcitedefaultseppunct}\relax
\EndOfBibitem
\bibitem{Sjostrand:2006za}
T.~Sj\"{o}strand, S.~Mrenna, and P.~Skands,
  \ifthenelse{\boolean{articletitles}}{{\it {PYTHIA 6.4 physics and manual}},
  }{}\href{http://dx.doi.org/10.1088/1126-6708/2006/05/026}{JHEP {\bf 05}
  (2006) 026}, \href{http://arxiv.org/abs/hep-ph/0603175}{{\tt
  arXiv:hep-ph/0603175}}\relax
\mciteBstWouldAddEndPuncttrue
\mciteSetBstMidEndSepPunct{\mcitedefaultmidpunct}
{\mcitedefaultendpunct}{\mcitedefaultseppunct}\relax
\EndOfBibitem
\bibitem{LHCb-PROC-2010-056}
I.~Belyaev {\em et~al.}, \ifthenelse{\boolean{articletitles}}{{\it {Handling of
  the generation of primary events in \gauss, the \lhcb simulation framework}},
  }{}\href{http://dx.doi.org/10.1109/NSSMIC.2010.5873949}{Nuclear Science
  Symposium Conference Record (NSS/MIC) {\bf IEEE} (2010) 1155}\relax
\mciteBstWouldAddEndPuncttrue
\mciteSetBstMidEndSepPunct{\mcitedefaultmidpunct}
{\mcitedefaultendpunct}{\mcitedefaultseppunct}\relax
\EndOfBibitem
\bibitem{Lange:2001uf}
D.~J. Lange, \ifthenelse{\boolean{articletitles}}{{\it {The EvtGen particle
  decay simulation package}},
  }{}\href{http://dx.doi.org/10.1016/S0168-9002(01)00089-4}{Nucl.\ Instrum.\
  Meth.\  {\bf A462} (2001) 152}\relax
\mciteBstWouldAddEndPuncttrue
\mciteSetBstMidEndSepPunct{\mcitedefaultmidpunct}
{\mcitedefaultendpunct}{\mcitedefaultseppunct}\relax
\EndOfBibitem
\bibitem{Allison:2006ve}
GEANT4 collaboration, J.~Allison {\em et~al.},
  \ifthenelse{\boolean{articletitles}}{{\it {Geant4 developments and
  applications}}, }{}\href{http://dx.doi.org/10.1109/TNS.2006.869826}{IEEE
  Trans.\ Nucl.\ Sci.\  {\bf 53} (2006) 270}\relax
\mciteBstWouldAddEndPuncttrue
\mciteSetBstMidEndSepPunct{\mcitedefaultmidpunct}
{\mcitedefaultendpunct}{\mcitedefaultseppunct}\relax
\EndOfBibitem
\bibitem{Agostinelli:2002hh}
GEANT4 collaboration, S.~Agostinelli {\em et~al.},
  \ifthenelse{\boolean{articletitles}}{{\it {GEANT4: A simulation toolkit}},
  }{}\href{http://dx.doi.org/10.1016/S0168-9002(03)01368-8}{Nucl.\ Instrum.\
  Meth.\  {\bf A506} (2003) 250}\relax
\mciteBstWouldAddEndPuncttrue
\mciteSetBstMidEndSepPunct{\mcitedefaultmidpunct}
{\mcitedefaultendpunct}{\mcitedefaultseppunct}\relax
\EndOfBibitem
\bibitem{LHCb-PROC-2011-006}
M.~Clemencic {\em et~al.}, \ifthenelse{\boolean{articletitles}}{{\it {The \lhcb
  simulation application, \gauss: design, evolution and experience}},
  }{}\href{http://dx.doi.org/10.1088/1742-6596/331/3/032023}{{J.\ of Phys.\
  Conf.\ Ser.\ } {\bf 331} (2011) 032023}\relax
\mciteBstWouldAddEndPuncttrue
\mciteSetBstMidEndSepPunct{\mcitedefaultmidpunct}
{\mcitedefaultendpunct}{\mcitedefaultseppunct}\relax
\EndOfBibitem
\bibitem{Breiman}
L.~Breiman, J.~H. Friedman, R.~A. Olshen, and C.~J. Stone, {\em Classification
  and regression trees}, Wadsworth international group, Belmont, California,
  USA, 1984\relax
\mciteBstWouldAddEndPuncttrue
\mciteSetBstMidEndSepPunct{\mcitedefaultmidpunct}
{\mcitedefaultendpunct}{\mcitedefaultseppunct}\relax
\EndOfBibitem
\bibitem{Roe}
B.~P. Roe {\em et~al.}, \ifthenelse{\boolean{articletitles}}{{\it {Boosted
  decision trees as an alternative to artificial neural networks for particle
  identification}},
  }{}\href{http://dx.doi.org/10.1016/j.nima.2004.12.018}{Nucl.\ Instrum.\
  Meth.\  {\bf A543} (2005) 577},
  \href{http://arxiv.org/abs/physics/0408124}{{\tt
  arXiv:physics/0408124}}\relax
\mciteBstWouldAddEndPuncttrue
\mciteSetBstMidEndSepPunct{\mcitedefaultmidpunct}
{\mcitedefaultendpunct}{\mcitedefaultseppunct}\relax
\EndOfBibitem
\bibitem{Hocker:2007ht}
A.~Hoecker {\em et~al.}, \ifthenelse{\boolean{articletitles}}{{\it {TMVA:
  Toolkit for Multivariate Data Analysis}}, }{}PoS {\bf ACAT} (2007) 040,
  \href{http://arxiv.org/abs/physics/0703039}{{\tt
  arXiv:physics/0703039}}\relax
\mciteBstWouldAddEndPuncttrue
\mciteSetBstMidEndSepPunct{\mcitedefaultmidpunct}
{\mcitedefaultendpunct}{\mcitedefaultseppunct}\relax
\EndOfBibitem
\bibitem{Pivk:2004ty}
M.~Pivk and F.~R. Le~Diberder, \ifthenelse{\boolean{articletitles}}{{\it
  {sPlot: a statistical tool to unfold data distributions}},
  }{}\href{http://dx.doi.org/10.1016/j.nima.2005.08.106}{Nucl.\ Instrum.\
  Meth.\  {\bf A555} (2005) 356},
  \href{http://arxiv.org/abs/physics/0402083}{{\tt
  arXiv:physics/0402083}}\relax
\mciteBstWouldAddEndPuncttrue
\mciteSetBstMidEndSepPunct{\mcitedefaultmidpunct}
{\mcitedefaultendpunct}{\mcitedefaultseppunct}\relax
\EndOfBibitem
\bibitem{CLsMethod}
A.~Read, \ifthenelse{\boolean{articletitles}}{{\it {Presentation of search
  results: the CL$_{\rm s}$ technique}},
  }{}\href{http://dx.doi.org/10.1088/0954-3899/28/10/313}{J.\ Phys.\  {\bf G28}
  (2002) 2693}\relax
\mciteBstWouldAddEndPuncttrue
\mciteSetBstMidEndSepPunct{\mcitedefaultmidpunct}
{\mcitedefaultendpunct}{\mcitedefaultseppunct}\relax
\EndOfBibitem
\bibitem{Junk:1999kv}
T.~Junk, \ifthenelse{\boolean{articletitles}}{{\it {Confidence level
  computation for combining searches with small statistics}},
  }{}\href{http://dx.doi.org/10.1016/S0168-9002(99)00498-2}{Nucl.\ Instrum.\
  Meth.\  {\bf A434} (1999) 435},
  \href{http://arxiv.org/abs/hep-ex/9902006}{{\tt arXiv:hep-ex/9902006}}\relax
\mciteBstWouldAddEndPuncttrue
\mciteSetBstMidEndSepPunct{\mcitedefaultmidpunct}
{\mcitedefaultendpunct}{\mcitedefaultseppunct}\relax
\EndOfBibitem
\end{mcitethebibliography}

\end{document}